\documentstyle[12pt,epsf]{article}
\def\hybrid{\topmargin -20pt    \oddsidemargin 0pt
        \headheight 0pt \headsep 0pt
        \textwidth 6.5in        
        \textheight 9in         
        \marginparwidth .875in
        \parskip 5pt plus 1pt   \jot = 1.5ex}

\hybrid
\makeatletter
\@addtoreset{equation}{section}
\makeatother

\newcommand{\cB}{{\cal B}}

\newcommand{\cM}{{\cal M}}

\newcommand{\cQ}{{\cal Q}}

\newcommand{\hf}{\frac12}
\newcommand{\qt}{\frac14}
\newcommand{\bea}{\begin{eqnarray}}
\newcommand{\eea}{\end{eqnarray}}
\newcommand{\be}{\begin{equation}}
\newcommand{\ee}{\end{equation}}
\newcommand{\bt}{\begin{tabular}}
\newcommand{\et}{\end{tabular}}
\newcommand{\ba}{\begin{array}}
\newcommand{\ea}{\end{array}}
\newcommand{\eps}{\epsilon}

\newcommand{\vev}[1]{\langle #1 \rangle}
\newcommand{\Tr}{\mathop{\rm Tr}}
\newcommand{\Pf}{\mathop{\rm Pf}}
\newcommand{\cof}{\mathop{\rm cof}}

\newcommand{\lra}{\leftrightarrow}

\def \msmall#1{{\scriptstyle{#1}}}

\def \half{{\textstyle\hf}}

\def \quarter{{\textstyle\qt}}
\def \soll={\stackrel{!}{=}}
\def \rf=#1{\stackrel{(\ref{#1})}{=}}
\def \bfm#1{\mbox{\boldmath$#1$}}

\def \Wtree{W_{\rm tree}}

\def \Wmag{W_{\rm mag}}

\def \w{{\rm with\ }}
\def \wh{{\rm where\ }}

\def \barfill{\leaders\hrule height 0.1 true pt\hfill}
\def \overbar#1{\vbox{\ialign{##\crcr\barfill\crcr\noalign{\kern 1pt
                                      \nointerlineskip}$\hfil{#1}\hfil$\crcr}}}
\def \scriptbar#1{{\vbox{\ialign{##\crcr\thinspace\barfill\thinspace\crcr
    \noalign{\kern 0.8pt\nointerlineskip}$\hfil{\scriptstyle #1}\hfil$\crcr}}}}

\newlength{\oldindent}
\newlength{\quadlength} 
\newlength{\abstand} \newlength{\breite}

\newcommand{\drawsquare}[2]{\hbox{%
\rule{#2pt}{#1pt}\hskip-#2pt
\rule{#1pt}{#2pt}\hskip-#1pt
\rule[#1pt]{#1pt}{#2pt}}\rule[#1pt]{#2pt}{#2pt}\hskip-#2pt
\rule{#2pt}{#1pt}}

\newcommand{\Yfun}{\raisebox{-.5pt}{\drawsquare{6.5}{0.4}}}
\newcommand{\Ysym}{\raisebox{-.5pt}{\drawsquare{6.5}{0.4}}\hskip-0.4pt%
        \raisebox{-.5pt}{\drawsquare{6.5}{0.4}}}
\newcommand{\Yasym}{\raisebox{-3.5pt}{\drawsquare{6.5}{0.4}}\hskip-6.9pt%
        \raisebox{3pt}{\drawsquare{6.5}{0.4}}}
\newcommand{\Yasthree}{\raisebox{-3.5pt}{\drawsquare{6.5}{0.4}}\hskip-6.9pt%
        \raisebox{3pt}{\drawsquare{6.5}{0.4}}\hskip-6.9pt
        \raisebox{9.5pt}{\drawsquare{6.5}{0.4}}}

\def \Yfunb{\overline{\Yfun}}
\def \Ysymb{\overline{\Ysym}}

\def \Yasymb{\overline{\Yasym}}

\def \Nucl#1{{\em Nucl.~Phys.}\ {\bf B#1}}
\def \NuclProc#1{{\em Nucl.~Phys.~Proc.~Suppl.}\ {\bf #1}}
\def \PhysR#1{{\em Phys.~Rev.}\ {\bf D#1}}
\def \PhysRL#1{{\em Phys.~Rev.~Lett.}\ {\bf #1}}
\def \PhysRev#1{{\em Phys.~Rev.}\ {\bf #1}}
\def \PhysRep#1{{\em Phys.~Rep.}\ {\bf #1}}
\def \PhysL#1{{\em Phys.~Lett.}\ {\bf #1B}}

\def \ModPL#1{{\em Mod.~Phys.~Lett.}\ {\bf A#1}}

\def \hep#1{{\tt hep-th/#1}}
\def \hepp#1{{\tt hep-ph/#1}}

\def \nonpert{non\discretionary{-}{}{-}per\-tur\-ba\-tive\ }

\def \nonan{non\discretionary{-}{}{-}anom\-a\-lous\ }
\def \nonab{non\discretionary{-}{}{-}Abel\-ian\ }
\def \nonvan{non\discretionary{-}{}{-}van\-ish\-ing\ }
\def \afun{antifundamental\ }
\def \asym{antisymmetric\ }

\renewcommand{\thefootnote}{\fnsymbol{footnote}}

\begin{document}

\begin{titlepage}
\begin{center}

\rightline{FTUAM-98/29}
\rightline{IFT-UAM/CSIC-98-29}
\rightline{hep-th/9812155}

\vskip .6in
{\LARGE \bf More Confining $N=1$ SUSY Gauge Theories\\[1ex]
From Non-Abelian Duality}
\vskip .8in

{\bf Matthias Klein\footnote{E-mail: Matthias.Klein@uam.es}
\\
\vskip 0.8cm
{\em Departamento de F\'\i sica Te\'orica C-XI and 
     Instituto de F\'\i sica Te\'orica C-XVI \\
     Universidad Aut\'onoma de Madrid, Cantoblanco,
     28049 Madrid, Spain}
}

\end{center}

\vskip 1.5cm

\begin{center} {\bf ABSTRACT } \end{center}
We expand on an idea of Seiberg that an $N=1$ supersymmetric gauge theory
shows confinement without breaking of chiral symmetry when the gauge symmetry
of its magnetic dual is completely broken by the Higgs effect. This has
recently been applied to some models involving tensor fields and an 
appropriate tree-level superpotential. We show how the confining spectrum of 
a supersymmetric gauge theory can easily be derived when a magnetic dual
is known and we determine it explicitly for many models containing fields
in second rank tensor representations. We also give the
form of the confining superpotential for most of these models.

\vfill
December 1998
\end{titlepage}

\renewcommand{\thefootnote}{\arabic{footnote}}
\setcounter{footnote}{0}

\section{Introduction}
There has been much progress in understanding the low-energy limit of
$N=1$ supersymmetric gauge theories during the last years. Due to 
holomorphy and non-renormalization theorems exact results could be  
obtained without the need to perform complicated calculations. As a
consequence it has become possible to argue that some gauge theories
with special matter content confine at low energies. The first example
is due to Seiberg \cite{Seib_D49} who found that supersymmetric quantum 
chromodynamics (SQCD) with gauge group $SU(N_c)$ and $N_f$ quark flavors 
(i.e.\ $N_f$ fields in the fundamental representation of the gauge group
and the same amount of fields in the \afun representation)
shows confinement when $N_f=N_c$ or $N_f=N_c+1$. In the former case
the chiral symmetry is spontaneously broken by a quantum deformation
of the classical moduli space, whereas in the latter case there exists
a point on the quantum moduli space where the full chiral symmetry is
unbroken. The low-energy theory is described by mesons and baryons
which, for $N_f=N_c+1$, are coupled by a non-perturbatively generated
superpotential.

This has been generalized to more complicated models. All $N=1$ 
supersymmetric gauge theories with vanishing tree-level superpotential 
which confine either with a quantum deformed moduli space and no 
non-perturbative superpotential or with a smooth confining superpotential
and without chiral symmetry breaking could be classified \cite{confine,qmm}
because they are constrained by an index argument.\footnote{There is another
class of confining gauge theories which do not have constraints for the
gauge invariant composite fields. They are classified in \cite{affineqm}.}
For the moduli space to 
be deformed by quantum effects the index\footnote{The relevance of this 
index was noticed by the authors of \cite{confine}. We take their 
normalization which assigns $\mu=1$ to the fundamental of
$SU$ and $Sp$ groups and $\mu=2$ to the vector of $SO$.}
$\Delta=\mu_{\rm matter}-\mu_G$ must vanish \cite{confine,qmm}, where 
$\mu_{\rm matter}$ denotes the sum over the Dynkin indices of all matter 
fields and $\mu_G$ is the Dynkin index of the adjoint representation. 
The condition $\Delta=2$ is necessary for the low-energy theory to be 
described by a superpotential which is a smooth function of the confined 
degrees of freedom \cite{confine} (this has been called {\em s-confinement} 
in \cite{confine}).

When a tree-level superpotential is present the index argument is no longer
valid. Because of the lower symmetry the non-perturbative superpotential is
less constrained and one expects more confining models to exist. Indeed,
Cs\'aki and Murayama \cite{newconf} recently showed that many of the 
Kutasov-like \cite{KuSch} models exhibit confinement (without breaking of 
chiral symmetry) for special values of the number of quark flavors $N_f$.%
\footnote{Some further confining models with \nonvan tree-level 
superpotential are discussed in \cite{TaYo}.} 
Fortunately, these theories simplify considerably once an appropriate 
superpotential for the tensor fields is added. More precisely, for all of 
the models considered in \cite{newconf} a dual description in terms of
magnetic variables is known \cite{KuSch,Intri,LS,ILSdual} and the authors of 
\cite{newconf} used the fact that the electric gauge theory confines when
its magnetic dual is completely higgsed.

Seiberg already used this idea as an additional consistency check in his 
original paper establishing electric-magnetic duality for \nonab 
$N=1$ supersymmetric
gauge theories \cite{Seib}. He showed how the confining superpotential
of SQCD with $N_f=N_c+1$ could be obtained by a perturbative calculation 
in the completely broken magnetic gauge theory. Under duality the 
fields of the magnetic theory (which are gauge singlets as the gauge 
symmetry is completely broken) are mapped to the mesons and baryons of 
the electric theory and the confining superpotential is easily shown 
to be the image of the magnetic superpotential under this mapping \cite{Seib}.
This a realization in $N=1$ supersymmetric gauge theories of an old idea of
't Hooft and Mandelstam \cite{dualmeiss} that confinement is driven by 
condensation of magnetic monopoles.

Now, many gauge theory models have been found that possess a dual description
in terms of magnetic variables in the infrared. This allows us to predict 
many new examples of confining gauge theories. The idea described in the
previous paragraph was first used by the authors of \cite{KSS} to determine 
the confining spectrum of the model proposed by Kutasov \cite{KuSch} and has 
recently been applied by Cs\'aki and Murayama \cite{newconf} to six further 
models that confine in the presence of an appropriate superpotential.

In the next section we explain how the duality discovered by Seiberg is
used to obtain the low-energy spectrum and the form of the \nonpert
superpotential of confining gauge theories. To this end we review the
example of SQCD with an additional field in the adjoint representation
discussed in \cite{KSS}. We then generalize the argument of 
\cite{confine} determining the most general \nonpert superpotential for
models with vanishing tree-level superpotential. We find that when 
tree-level terms are present the \nonpert superpotential is still
constrained but no longer uniquely determined by the symmetries alone.

In sections 3 and 4 we present the new confining gauge theories. We consider 
all gauge theory models of \cite{ILSdual,patterns} based on simple gauge 
groups and find that each of them confines when the gauge group of its 
magnetic dual gets completely broken by the Higgs effect. Six of these
models have recently been found to show confinement by the authors of 
\cite{newconf}. For nine of these theories, however, the confining phase
has not yet been discussed. We explain how the confining spectrum can easily 
be obtained from the duality mapping and determine the form of the confining
superpotential for most of the models. Constructing the confined low-energy
spectrum for theories with $SU(N_c)$ or $SO(N_c)$ gauge groups involves
considering generalized baryons that can be mapped to similar operators
of the magnetic theory for any value of $N_c$. When $N_c$ is tuned such
that the magnetic gauge group is completely broken these mappings reduce
to a correspondence between the light degrees of freedom of the magnetic
theory in the Higgs phase and the confined degrees of freedom of the
electric theory in the confinement phase. Furthermore, some of the terms
of the confining superpotential can be obtained by applying these mappings
to the tree-level superpotential of the magnetic theory.
 
It is interesting to deform these theories either by giving a large
expectation value to one of the composite operators or by adding a mass
term for one quark flavor and integrating out the massive modes. 
In the first case one flows to a theory which is again confining with 
a smooth \nonpert superpotential.\footnote{There may be classical flat 
directions which lead to effective theories with no stable ground state. 
These flat directions are removed from the moduli space by quantum effects 
and therefore do not correspond to composite operators of the confined 
low-energy spectrum \cite{newconf}.}
The mass deformation leads to a theory with one quark flavor less, which in
most of the cases of section 3 has no stable vacuum. However, we
find that for $SU(N_c)$ with an \asym flavor and for $Sp(2N_c)$ with an
\asym tensor there exists a quantum moduli space with multiple constraints
on the composite operators, one of which is modified quantum mechanically.
This is, to our knowledge, the first example of theories that possess a 
quantum modified moduli space in the presence of a tree-level superpotential.

\section{Confining gauge theories from \nonab duality}
It is an interesting fact that the s-confining models with vanishing 
tree-level superpotential found in \cite{confine} can be derived 
from the completely higgsed magnetic dual whenever a duality is known
for the model that contains one quark flavor more. Let us make two remarks:
(i) There are 25 s-confining models with $Spin(N_c)$ gauge groups, 
$N_c$ ranging from 7 to 14, which we do not consider here. Nine of these 
models can be derived from a self-dual description of the theory with one 
more vector as explained in \cite{CSST} and seven further ones can presumably 
be derived from the self-dual $Spin(N_c)$ models presented in \cite{Karch}.
(ii) The confining phase of an $SO(N_c)$ gauge theory with $N_f=N_c-3$
quarks, as discussed in \cite{Seib,SOdual}, can also be derived from the dual
magnetic $SO(N_f-N_c+4)$ gauge theory. But this model is not s-confining
in the sense of \cite{confine} because its quantum moduli space possesses
two distinct branches and confinement is found only on one of these
branches.

\begin{table}[t]
$$\ba{|c|c|c|c|c|}
\hline\rule{0mm}{2.5ex}
{\rm electric} &{\rm microscopic} &{\rm duality} &{\rm magnetic} &{\rm ref.}\\
\rule[-1ex]{0mm}{1ex}
{\rm gauge\ group} &{\rm spectrum} &{\rm known\ for\ \ldots} 
&{\rm gauge\ group} &\ \\ \hline
\rule{0mm}{3ex} SU(N_c) &(N_c+1)\,(\Yfun+\Yfunb) &N_f\,(\Yfun+\Yfunb) 
                &SU(N_f-N_c) &\cite{Seib}\\
\rule{0mm}{3ex}SU(N_c) &\Yasym+\Yasymb+3\,(\Yfun+\Yfunb) 
               &\Yasym+\Yasymb+4\,(\Yfun+\Yfunb) &SU(N_c) &\cite{CSST}\\
\rule{0mm}{3ex}SU(N_c) &\Yasym+4\,\Yfun+N_c\,\Yfunb 
               &\Yasym+N_f\,\Yfun &SU(N_f-3) &\\
               &&+(N_c+N_f-4)\,\Yfunb &\ \times Sp(2(N_f-4)) &\cite{Pou2}\\
\rule{0mm}{3ex} SU(7) &2\,(\Yasym+3\Yfunb) &- &- &\ \\
\rule{0mm}{4ex} SU(6) &\Yasthree+4\,(\Yfun+\Yfunb) 
                &\Yasthree+5\,(\Yfun+\Yfunb) &Sp(4) &\cite{CSST}\\
\rule{0mm}{3ex} SU(6) &2\,\Yasym+\Yfun+5\,\Yfunb &- &- &\ \\
\rule{0mm}{3ex} SU(5) &3\,(\Yasym+\Yfunb) &- &- &\ \\
\rule[-1.5ex]{0mm}{4.5ex} SU(5) &\Yasym+2\,\Yfun+4\,\Yfunb &- &- &\ \\
\hline
\rule{0mm}{3ex} Sp(2N_c) &2\,(N_c+2)\,\Yfun &2\,N_f\,\Yfun 
                &Sp(2(N_f-\!N_c-\!2)) &\cite{IP}\\
\rule[-1.5ex]{0mm}{4.5ex} Sp(2N_c) &\Yasym+6\,\Yfun &\Yasym+8\,\Yfun &Sp(2N_c) 
                      &\cite{CSS}\\
\hline
\rule[-1ex]{0mm}{4ex} G_2 &5\,\Yfun &- &- & \cite{G2}\\ \hline
\ea$$
\caption{\label{sconfmodels}All s-confining models of \cite{confine} except 
those containing spinors. For some of them a magnetic dual is known. The 
confining phase can in these cases be derived by completely breaking the 
magnetic gauge group.}
\end{table}

In table \ref{sconfmodels} we list all s-confining models of \cite{confine} 
which do not contain spinors. In the first two columns we give the gauge 
group\footnote{In this paper $Sp(2N_c)$ denotes the symplectic group of rank 
$N_c$.} and the microscopic spectrum of the electric theory that leads to 
confinement. The third column displays the spectrum of the electric theory 
for which a dual magnetic description is known. The magnetic gauge group is 
shown in the fourth column. Finally, for each model the reference is given 
where the mentioned duality is discussed and where in most cases it is 
shown how to obtain the s-confining model from the magnetic dual.%
\footnote{ The $G_2$ gauge theory was first treated in \cite{G2}. The 
suggestion of \cite{Karch} that the theory with six fundamental fields 
might be self-dual is not well established because the 't Hooft anomalies 
for the discrete symmetries do not match between the electric and the 
magnetic theory \cite{discranom}.}

Let us explain the idea of using duality to obtain the confining spectrum
by briefly reviewing the example worked out in \cite{KSS}. Consider SQCD
with an additional second rank tensor field $X$ transforming in the adjoint 
representation. Under the gauge and global symmetries the matter fields
transform like in the following table. (The charges are chosen such that all 
these symmetries are non-anomalous.\footnote{Anomalies of discrete symmetries 
are discussed in \cite{discranom,IR,BD}})
$$\ba{|c|c|ccccc|}
\hline  \rule[-1.3ex]{0mm}{4ex}
& SU(N_c) & SU(N_f)_L & SU(N_f)_R & U(1)_B 
& U(1)_R & {\bfm Z}_{(k+1)N_f} \\ \hline
\rule{0mm}{3ex}
Q  &\Yfun &\Yfun &\bfm 1 &{1\over N_c} &1-{2N_c\over(k+1)N_f}& -N_c\\
\bar Q  &\Yfunb &\bfm 1  &\Yfun &-{1\over N_c} &1-{2N_c\over(k+1)N_f} &-N_c\\ 
\rule[-1.5ex]{0mm}{1.5ex}
X  &adj &\bfm 1 &\bfm 1 &0 &{2\over k+1} &N_f\\ \hline 
\ea$$

The low-energy limit of the theory with vanishing tree-level superpotential 
is not yet understood. However the situation turned out to simplify 
\cite{KuSch} when a superpotential term
\be \label{Weladj} W=h\,\Tr X^{k+1}
\ee
is added. Here $k$ denotes a positive integer and $h$ is a dimensionful
coupling parameter.

This model has a number of flat directions which can be parametrized 
by the expectation values of gauge invariant composite operators.
These operators are most conveniently written in terms of ``dressed'' 
quarks $Q_{(j)}\equiv X^j Q$, $\bar Q_{(j)}\equiv X^j \bar Q$, where
the gauge indices are contracted with a Kronecker delta, \cite{KuSch}:
\bea \label{mesbaradj}
     mesons  &&M_j\equiv Q Q_{(j)},\quad j=0,\ldots,k-1,\nonumber\\
     baryons &&B^{(n_0,\ldots,n_{k-1})}\equiv (Q)^{n_0}(Q_{(1)})^{n_1}
               \cdots(Q_{(k-1)})^{n_{k-1}}, \quad\sum_{j=0}^{k-1}n_j=N_c,\\
\hbox{\em antibaryons} &&\bar B^{(\bar n_0,\ldots,\bar n_{k-1})}\equiv 
               (\bar Q)^{\bar n_0}(\bar Q_{(1)})^{\bar n_1}
               \cdots(\bar Q_{(k-1)})^{\bar n_{k-1}}, 
               \quad\sum_{j=0}^{k-1}\bar n_j=N_c,\nonumber
\eea
where the gauge indices are contracted with a Kronecker delta for the mesons
and an epsilon tensor for the (anti-)baryons. The classical flat directions 
corresponding to the operators $\Tr X^j$, $j=2,\ldots k$, in general are 
lifted by the superpotential (\ref{Weladj}). For vanishing quark expectation
values $\vev X$ must vanish as well if $N_c$ is no multiple of $k$.

The infrared behavior of this theory can equivalently be described by 
a magnetic $SU(\tilde N_c)$ gauge theory, with $\tilde N_c=kN_f-N_c$ 
and matter content \cite{KuSch}
$$\ba{|c|c|ccccc|}
\hline  \rule[-1.3ex]{0mm}{4ex}
& SU(\tilde N_c) & SU(N_f)_L & SU(N_f)_R & U(1)_B 
& U(1)_R & {\bfm Z}_{(k+1)N_f} \\ \hline
\rule{0mm}{3ex}
q  &\Yfun &\Yfunb &\bfm 1 &{1\over\tilde Nc} 
          &1-{2\tilde N_c\over(k+1)N_f}& -\tilde N_c\\
\bar q  &\Yfunb &\bfm 1 &\Yfunb &-{1\over\tilde Nc} 
                &1-{2\tilde N_c\over(k+1)N_f} &-\tilde N_c \\ 
Y  &adj &\bfm 1 &\bfm 1 &0 &{2\over k+1} &N_f\\ 
\rule[-1.5ex]{0mm}{1.5ex}
M_j  &\bfm 1 &\Yfun &\Yfun &0 &2-{4N_c-2jN_f\over(k+1)N_f} 
             &-2N_c+jN_f\\ \hline
\ea$$
The magnetic theory contains a tree-level superpotential
\be \label{Wmagadj}
    \Wmag=-h\,\Tr Y^{k+1}\ + \ {h\over\mu^2}\,\sum_{j=0}^{k-1}
                                M_{k-1-j}qY^j\bar q,
\ee
where $\mu$ is a mass scale that has to be introduced to match the
magnetic to the electric theory.

Under duality the gauge singlets $M_j$ of the magnetic theory are mapped
to the meson operators of (\ref{mesbaradj}) and are therefore denoted
by the same symbols. The correct degrees of freedom $\hat M_j$ of the 
magnetic theory have mass dimension one and are related to the operators 
$M_j$ by $\mu^{j+1}\hat M_j=M_j$. We prefer to express all equations
in terms of the $M_j$. One can construct magnetic baryon operators
$\widetilde B^{(m_0,\ldots,m_{k-1})}$, 
$\widetilde{\bar B}{}^{(\bar m_0,\ldots,\bar m_{k-1})}$ 
by contracting products of $\tilde N_c$ dressed magnetic quarks with an 
epsilon tensor very much like in the electric theory. The mapping to the 
electric baryons of (\ref{mesbaradj}) is given by \cite{KuSch}
\be \label{barmapadj}
B^{(n_0,\ldots,n_{k-1})}\lra \widetilde B^{(m_0,\ldots,m_{k-1})},\quad
\w \ m_j=N_f-n_{k-1-j}.
\ee

Now consider the model with $N_f+1$ quark flavors and choose
$N_c=kN_f-1$ \cite{KSS,newconf}. Adding a mass term for the $(N_f+1)$th 
flavor in the electric theory and integrating out the massive modes leads 
to a complete breaking of the $SU(k+1)$ gauge symmetry on the magnetic 
side via the Higgs effect (in general the number of magnetic colors 
is reduced by $k$ when integrating out a flavor from the electric theory).
One color component of each quark flavor stays massless after the symmetry
breaking. These singlets can be identified with the magnetic baryons 
$\widetilde B^{(1,0,\ldots,0)}$, $\widetilde{\bar B}{}^{(1,0,\ldots,0)}$ 
and via (\ref{barmapadj}) are mapped to the electric baryons 
$B^{(N_f,\ldots,N_f,N_f-1)}$, 
$\bar B^{(N_f,\ldots,N_f,N_f-1)}$ of (\ref{mesbaradj}). At low energies
the magnetic theory is in the weakly coupled Higgs phase and the electric
theory is very strongly coupled and does not flow to a fixed point of
the remormalization group.\footnote{An infrared fixed point is only
expected in the range of parameters where both the electric and the
magnetic theory are asymptotically free in the ultraviolet 
\cite{Seib,KuSch}.} Thus, if there still exists a sensible 
description of the low-energy theory in terms of electric variables it 
should only consist of the confined degrees of freedom. Indeed, one finds
\cite{KSS,newconf} that the electric theory confines with low-energy
spectrum given by the mesons $M_j$, $j=0,\ldots,k-1$, and the baryons
$B^{(N_f,\ldots,N_f,N_f-1)}$, $\bar B^{(N_f,\ldots,N_f,N_f-1)}$ of 
(\ref{mesbaradj}). The confining superpotential has been derived in 
\cite{newconf} and is exactly reproduced by the effective superpotential 
of the magnetic theory when care is taken of instanton effects in the 
completely broken magnetic gauge group.

Integrating out one quark flavor from the $N_c=kN_f-1$ theory the authors
of \cite{newconf} found a \nonpert superpotential of 
Affleck--Dine--Seiberg \cite{ADS} type. This was to be expected from the
analysis of \cite{KuSch} which showed that the theory has no stable
vacuum for $N_f<N_c/k$. However, this behavior is in contrast to the
s-confining gauge theories without a tree-level superpotential. 
Integrating out one flavor from these theories one obtains a quantum 
modified moduli space with no \nonpert superpotential.

In the following we want to use this Higgs phase / confinement duality
to determine the confining spectrum of many new models. One just has to 
find the mapping of the magnetic singlets that stay massless after higgsing 
the gauge group to the gauge invariant composite fields of the electric 
theory. A straightforward way to find this mapping for models with 
$SU$ or $SO$ gauge groups is to consider baryonic composite operators
of the electric theory and the duality mapping to their magnetic counterparts.
When the magnetic gauge group is higgsed the light degrees of freedom are
proportional to some generalized baryons and can therefore easily be 
mapped to the corresponding confined degrees of freedom of the electric
theory. Some of the terms of the confining superpotential are obtained by 
applying this mapping to the magnetic tree-level superpotential others 
are generated by instanton effects in the completely broken magnetic 
gauge group. However, in practice it is very difficult to determine
the precise form of the superpotential terms that are generated by
instantons of the magnetic gauge group. It is therefore important
to know the \nonpert electric superpotential from independent 
considerations.

For any $N=1$ supersymmetric model with vanishing tree-level superpotential
the form of the most general superpotential that can be generated by \nonpert 
effects is completely fixed by the requirement that it be invariant under all 
symmetries of the considered model \cite{ADS,ILS,confine}. For a theory
with gauge group $G$ and chiral matter fields $\phi_l$ in representations
$r_l$ of $G$ and dynamically generated scale $\Lambda$ one finds
\be \label{Weff}
W \propto \left({\prod_l(\phi_l)^{\mu_l}\over\Lambda^b}\right)^{2\over\Delta},
\ee
where $\mu_l$ is the (quadratic) Dynkin index of the representation $r_l$,
$\mu_G$ denotes the index of the adjoint representation, 
$\Delta=\sum_l\mu_l-\mu_G$ and $b=\half(3\mu_G-\sum_l\mu_l)$ is the coefficient
of the 1-loop $\beta$-function. In general the complete \nonpert
superpotential consists of a sum of terms of the form (\ref{Weff})
with different possible contractions of all gauge and flavor indices.
The relative coefficients of these terms cannot be fixed by symmetry
arguments but must be inferred from a different reasoning.

When a tree-level superpotential is present the global symmetries are 
reduced and hence the non-perturbative corrections less constrained.
To see this more explicitly divide the set of matter fields into two
subsets $\{\phi_l\}=\{\bar\phi_{\bar l}\}\cup\{\hat\phi_{\hat l}\}$,
with $\{\bar\phi_{\bar l}\}\cap\{\hat\phi_{\hat l}\}=\emptyset$,
and add one tree-level term for the hatted fields:
\be \label{Wtreehat}
\Wtree=h\,\prod_{\hat l}\left(\hat\phi_{\hat l}\right)^{n_{\hat l}},
\ee
where $h$ is a dimensionful coupling parameter and the $n_{\hat l}$
are positive integers.

The form of the \nonpert superpotential can be easily derived by viewing
the dynamical scale $\Lambda$ and the coupling parameter $h$ as background 
chiral fields \cite{Seib_L318}. For each field $\phi_l$ there is classically 
a $U(1)_l$ symmetry under which only $\phi_l$ carries charge 1 and all other 
fields carry charge 0. The $U(1)_{\hat l}$ symmetries are spontaneously 
broken by the tree-level term (\ref{Wtreehat}) but can be restored by assigning
charge $-n_{\hat l}$ under each of them to the background field $h$.
At the quantum level these symmetries are anomalous but this can be cured
by assigning charge $\mu_l$ to $\Lambda^b$ \cite{ILS,ISlectures}. Requiring 
that the \nonpert superpotential be invariant under all these symmetries and 
have charge 2 under the \nonan R-symmetry we find
\be \label{Weff2}
W \propto\left({\prod_{\bar l}(\bar\phi_{\bar l})^{\mu_{\bar l}}\over\Lambda^b}
          \right)^\alpha\ \prod_{\hat l}\left(\hat\phi_{\hat l}
          \right)^{\beta_{\hat l}}\ h^\gamma,
\ee
where the powers $\alpha$, $\beta_{\hat l}$, $\gamma$ must verify the
following relations:
\bea \label{albega}
\gamma &= &1-\half\alpha\Delta,\nonumber \\
\beta_{\hat l} &= &\mu_{\hat l}\,\alpha+\gamma n_{\hat l}.
\eea
We conclude that the symmetries do not uniquely fix the form of the
\nonpert superpotential. For a given theory, i.e.\ fixed $\Delta$, 
there exists a superpotential consistent with all symmetries for each
value of the power $\alpha$. However in some cases it is possible to
determine the allowed values for $\alpha$ from a different reasoning, 
as we shall see later. Note that for $\alpha=0$ we recover the tree-level
superpotential. In the following we will be interested in the cases where
$W$ is a smooth function of the $\phi_l$, i.e.\ $\alpha=k$, $k$ a positive 
integer. Such a superpotential term can be generated by a $k$-instanton 
effect in the completely broken magnetic gauge group. This is because
the dynamically generated scale $\tilde\Lambda$ of the magnetic theory is
related to the electric scale $\Lambda$ by an equation of the form 
$\Lambda^b\tilde\Lambda^{\tilde b}=f(h,\mu),$ where $h$ is the coupling 
of (\ref{Wtreehat}), $\mu$ is a mass scale similar to the one introduced
in (\ref{Wmagadj}) and $f$ is some function of $h$ and $\mu$ such that
the equation is invariant under all $U(1)_l$-symmetries. (For all of
the models considered in section 3 this function is given by 
$f(h,\mu)=\mu^{N_f+\bar N_f}h^{-\Delta}$, where $N_f$ ($\bar N_f$) is the
number of (anti-)fundamental fields.) Thus the magnetic
dual of (\ref{Weff2}) is proportional to $\tilde\Lambda^{\tilde b\alpha}$
and can be generated by an $\alpha$-instanton effect. 
In the limit of vanishing tree-level superpotential, which can formally be
obtained by setting $\gamma=0$, we find $\alpha=2/\Delta$ because of
(\ref{albega}) and therefore recover the index argument of \cite{confine}
that $\Delta$ has to equal 2 when one demands $W$ to be smooth in the fields.

Next, let us generalize this to the case of two tree-level terms 
\be \label{Wtreehat2}
\Wtree=h_1\,\prod_{\hat l\in S_1}\left(\hat\phi_{\hat l}\right)^{n_{\hat l}}
     +h_2\,\prod_{\hat l\in S_2}\left(\hat\phi_{\hat l}\right)^{m_{\hat l}},
\ee
where $S_1$, $S_2$ are two subsets of $\{\hat l\}$ such that 
$S_1\cup S_2=\{\hat l\}$. A similar reasoning leads to a \nonpert
superpotential of the form (\ref{Weff2}) with $h^\gamma$ replaced
by $h_1^{\gamma_1}h_2^{\gamma_2}$ and (\ref{albega}) replaced by
\bea \label{albega2}
\gamma_1+\gamma_2 &= &1-\half\alpha\Delta,\nonumber \\
\beta_{\hat l} &= &\mu_{\hat l}\,\alpha+\eps_1\gamma_1 n_{\hat l}
                                       +\eps_2\gamma_2 m_{\hat l},\\
{\rm where} && \eps_{1/2}=\left\{ \ba{ll} 0 &{\rm if}\ \hat l\not\in S_{1/2}\\
                                   1 &{\rm if}\ \hat l\in S_{1/2}\ea\right. .
\nonumber
\eea
We see that the \nonpert superpotential is even less constrained. For a
given theory the correct form of the superpotential can only be determined
if we know $\alpha$ and one of the $\gamma$'s from different arguments.
Note that in the limit $S_2=\emptyset$ and $\gamma_2=0$ we recover the case
of only one tree-level term (\ref{Weff2}, \ref{albega}).
These considerations are only valid if the supersymmetric field strength
$W_\alpha$ does not appear in the superpotential. One could however imagine
that the superpotential (\ref{Weff2}) is multiplied by $(W_\alpha)^{2\delta}$,
where $\delta$ is a positive integer. $W_\alpha$ has R-charge 1 but does not 
carry charge under any of the $U(1)_l$ symmetries and therefore only the 
relation for $\gamma$ (first line in (\ref{albega}) and (\ref{albega2})) 
is modified:
\be \label{albega3}
\gamma_1+\gamma_2 = 1-\half\alpha\Delta-\delta.
\ee
(The case of only one tree-level term is obtained by setting $\gamma_2=0$ and
$\gamma=\gamma_1$.)

Additional information on the confining superpotential comes from the fact
that the equations of motion derived from this superpotential should reproduce
the classical constraints that hold amongst the confined degrees of freedom.
In some cases it is easy to determine at least some of these constraints.
The superpotential terms necessary to produce them can then be constructed.
The authors of \cite{newconf} found a simple method to determine explicitly
the constraints on the gauge invariant composite operators which works for
some gauge theories with only adjoint and fundamental matter. In
these cases all the confined degrees of freedom can be expressed only in terms
of the dressed quarks introduced in the line above  eq.\ (\ref{mesbaradj}). 
To derive the constraints the considered theory can therefore be viewed as 
containing no tensor field but having an enhanced number of quark flavors. 
In the above example of $SU(N_c)$ with an adjoint tensor this means that we 
treat it as a theory of $kN_f=N_c+1$ quark flavors. However, the constraints
amongst the meson and baryon operators in SQCD with one more flavor than
colors are known \cite{Seib_D49}.

Most of the models containing tensor fields that show the phenomenon of
\nonab duality in the presence of an appropriate tree-level 
superpotential have been presented in a systematic way by the authors of
\cite{patterns}. They realized that the superpotentials for the tensor
fields in these models resemble $A_k$ or $D_k$ singularity forms,
generalized from numbers to matrices. The models containing one tensor
(and its conjugate for complex representations) have superpotentials of
the form $\Tr X^{k+1}$, corresponding to an $A_k$ singularity under the
usual ADE classification \cite{ADEsing}. The models containing two tensors
(and their conjugates for complex representations) have superpotentials
of the form $\Tr X^{k+1}+\Tr XY^2$, corresponding to a $D_{k+2}$
singularity. We find that all the models of \cite{patterns} based on
simple gauge groups confine when the magnetic gauge group is completely
higgsed (for most of the $A_k$ models this has already been established
in \cite{newconf}). Using the methods described above we determine the
low-energy spectrum in each case and in most cases also the form of 
the confining superpotential. An overview of the results is given in table 
\ref{results}. It shows for each model (specified by the gauge group, matter 
content and tree-level superpotential) the number of colors for which the 
model confines and the powers $\alpha$, $\beta$ and $\gamma$ of the \nonpert
superpotential (\ref{Weff2}). For some models there are terms with different
values of these coefficients. In these cases we display the powers that
correspond to the terms with  highest $\alpha$ because only these terms
are relevant for deriving the classical constraints. For the $D_{k+2}$ models 
we restrict ourselves to give the value of $\alpha$. The powers
$\beta$ and $\gamma$ could not be uniquely fixed. For some superpotential
terms they are calculated in section 4.

\begin{table}[h]
\vskip-5mm
$$\ba{|c|c|c|c|c|} \hline 
\rule{0mm}{3ex} &\multicolumn{4}{c|}{SU(N_c)}\\ \hline
\rule{0mm}{3ex} {\rm tensors} &adj &\Yasym+\Yasymb &\Ysym+\Ysymb 
                              &\Yasym+\Ysymb\\
\rule{0mm}{3ex} \Wtree &X^{k+1} &(X\bar X)^{k+1} &(X\bar X)^{k+1} 
                       &(X\bar X)^{2(k+1)}\\
\rule{0mm}{3ex} N_c &kN_f-1 &(2k+1)N_f-4k-1 &(2k+1)N_f+4k-1 &(4k+3)(N_f+4)-1\\
\rule{0mm}{3ex} \alpha &k &1 &2(k+1) &2(k+1) \\
\rule{0mm}{3ex} \beta_X &(k-1)N_c &k(N_f-1) &(k+1)(2k(N_f+2)-1) 
                &\msmall{2(k+1)((2k+1)(N_f+4)-3)}\\
\rule{0mm}{3ex} \beta_{\bar X} &  &k(N_f-1) &(k+1)(2k(N_f+2)-1) 
                &\msmall{2(k+1)((2k+1)(N_f+4)+1)}\\
\rule{0mm}{3ex} \gamma &-N_c &3-N_f &-(N_c+N_f+4) &-2(k+1)(N_f+4)\\ \hline
\ea$$
$$\ba{|c|c|c|c|c|} \hline \rule{0mm}{3ex} &\multicolumn{2}{c}{Sp(2N_c)} 
                        &\multicolumn{2}{|c|}{SO(N_c)}\\ \hline
\rule{0mm}{3ex} {\rm tensors} &\Ysym &\Yasym &\Yasym &\Ysym\\
\rule{0mm}{3ex} \Wtree &X^{2(k+1)} &X^{k+1} &X^{2(k+1)} &X^{k+1}\\
\rule{0mm}{3ex} N_c &(2k+1)N_f-2 &k(N_f-2) &(2k+1)N_f+3 &k(N_f+4)-1\\
\rule{0mm}{3ex} \alpha &2k+1 &1 &1 &k\\
\rule{0mm}{3ex} \beta &2k(N_c+1) &(k-1)(N_f-1) &2k(N_f+1)+4 &(k-1)(N_c-2k)\\
\rule{0mm}{3ex} \gamma &-(N_c+1) &3-N_f &1-N_f &-(N_c+2k)\\ \hline
\ea$$

\vskip2ex
$$\ba{|c|c|c|c|c|} \hline 
\rule{0mm}{3ex} &\multicolumn{4}{c|}{SU(N_c)}\\ \hline
\rule{0mm}{3ex} {\rm tensors} &2\,adj &adj+\Yasym+\Yasymb &adj+\Ysym+\Ysymb 
                              &adj+\Yasym+\Ysymb\\
\rule{0mm}{3ex} \Wtree &X^{k+1}+XY^2 &X^{k+1}+XY\bar Y &X^{k+1}+XY\bar Y 
                       &X^{k+1}+XY\bar Y\\
\rule{0mm}{3ex} N_c &3kN_f-1 &3kN_f-5 &3kN_f+3 &3k(N_f+4)-1\\
\rule{0mm}{3ex} \alpha &3k &k &4k &2k \\ \hline
\ea$$
$$\ba{|c|c|c|c|c|} \hline \rule{0mm}{3ex} &\multicolumn{2}{c}{Sp(2N_c)} 
                        &\multicolumn{2}{|c|}{SO(N_c)}\\ \hline
\rule{0mm}{3ex} {\rm tensors} &2\,\Yasym &\Yasym+\Ysym &2\,\Ysym 
                              &\Ysym+\Yasym\\
\rule{0mm}{3ex} \Wtree &X^{k+1}+XY^2 &X^{k+1}+XY^2 &X^{k+1}+XY^2 
                       &X^{k+1}+XY^2\\
\rule{0mm}{3ex} N_c &3kN_f-4k-2 &3kN_f-4k+2 &3kN_f+8k+3 &3kN_f+8k-5\\
\rule{0mm}{3ex} \alpha &(k) &(3k) &3k &k\\ \hline
\ea$$
\caption{\label{results}Gauge theories that confine in the presence of a
  tree-level superpotential. The coefficients $\alpha$, $\beta$, $\gamma$
  refer to the powers in the \nonpert superpotential (\ref{Weff2}). Some of
  the terms of this superpotential are generated by an $\alpha$-instanton 
  effect in the dual magnetic gauge theory. For $Sp(2N_c)$ with two tensors
  we suppose that $\alpha$ has the given values although we were not able 
  to prove this.}
\end{table}


\section{Models with $A_k$-type superpotentials}
\subsection{$SU(N_c)$ with an \asym tensor and its conjugate}
Consider SQCD with an additional flavor of \asym tensors $X$, $\bar X$
and tree-level superpotential $\Wtree=h\,\Tr(X\bar X)^{k+1}$. 
This model was first studied in \cite{ILSdual}. The transformation
properties of the matter fields under the gauge symmetry and the \nonan
global symmetries are shown in the following table:
$$\ba{|c|c|cccccc|}
\hline   \rule[-1.3ex]{0mm}{4ex}
& SU(N_c) & SU(N_f)_L & SU(N_f)_R &U(1)_X & U(1)_B 
& U(1)_R & {\bfm Z}_{2(k+1)N_f} \\ \hline
\rule{0mm}{3ex}
Q  &\Yfun &\Yfun &\bfm 1 &0 &{1\over N_c} &1-{N_c+2k\over(k+1)N_f} &-(N_c-2)\\
\bar Q  &\Yfunb &\bfm 1 &\Yfun &0 &-{1\over N_c} 
                &1-{N_c+2k\over(k+1)N_f} &-(N_c-2)\\ 
X  &\Yasym &\bfm 1 &\bfm 1 &1 &{2\over N_c} &{1\over k+1} &N_f\\
\rule[-1.5ex]{0mm}{1.5ex}
\bar X  &\Yasymb &\bfm 1 &\bfm 1 &-1 &-{2\over N_c} &{1\over k+1} &N_f\\ 
\hline 
\ea$$

We will be interested in the following gauge invariant composite operators 
that can be built from the elementary fields:
\bea \label{mesbaras}
mesons && M_j=Q\bar Q_{(j)},\quad
   P_r= Q\bar XQ_{(r)},\ \bar P_r=\bar QX\bar Q_{(r)},\nonumber\\
   &&\w Q_{(j)}=(X\bar X)^j Q,\ \bar Q_{(j)}=(\bar XX)^j\bar Q,\quad 
     j=0,\ldots,k,\quad r=0,\ldots,k-1,\nonumber \\
baryons && B_n=X^n Q^{N_c-2n},\quad \bar B_n=\bar X^n \bar Q^{N_c-2n},\quad
   n=0,\ldots,\left\lceil{N_c\over2}\right\rceil,\\
&& T_i=\Tr(X\bar X)^i,\quad i=1,\ldots,k,\nonumber
\eea
where the gauge indices of the baryons are contracted with an epsilon tensor.

This theory is dual to an $SU(\tilde N_c)$ gauge theory, with 
$\tilde N_c=(2k+1)N_f-4k-N_c$ and matter content \cite{ILSdual}
$$\ba{|c|c|cccccc|}
\hline   \rule[-1.3ex]{0mm}{4ex}
& SU(\tilde N_c) & SU(N_f)_L & SU(N_f)_R &U(1)_X & U(1)_B 
& U(1)_R & {\bfm Z}_{2(k+1)N_f} \\ \hline
\rule{0mm}{3ex}
q  &\Yfun &\Yfunb &\bfm 1 &{k(N_f-2)\over\tilde N_c} &{1\over\tilde N_c} 
          &1-{\tilde N_c+2k\over(k+1)N_f} &N_c+N_f-2\\
\bar q  &\Yfunb &\bfm 1 &\Yfunb &-{k(N_f-2)\over\tilde N_c} 
          &-{1\over\tilde N_c} &1-{\tilde N_c+2k\over(k+1)N_f} 
          &N_c+N_f-2\\ 
Y  &\Yasym &\bfm 1 &\bfm 1 &{N_c-N_f\over\tilde N_c} &{2\over\tilde N_c}
           &{1\over k+1} &N_f\\
\rule[-1.5ex]{0mm}{1.5ex}
\bar Y  &\Yasymb &\bfm 1 &\bfm 1 &-{N_c-N_f\over\tilde N_c} 
           &-{2\over\tilde N_c} &{1\over k+1} &N_f\\ \hline
\ea$$
and singlets $M_j$, $P_r$, $\bar P_r$ that carry the same quantum numbers
as the mesons of the electric theory.

The following tree-level superpotential of the magnetic theory is invariant
under all the symmetries:
\be \label{Wmagas}
   \Wmag =-h\,\Tr (Y\bar Y)^{k+1}
          +{h\over\mu^2}\,\sum_{j=0}^k M_{k-j} q\bar q_{(j)}
          +{h\over\mu^2}\,\sum_{r=0}^{k-1} \Big[P_{k-1-r} q\bar Y q_{(r)}
                    +\bar P_{k-1-r}\bar qY\bar q_{(r)}\Big]\ ,
\ee
where magnetic dressed quarks have been introduced by $q_{(j)}=(Y\bar Y)^j q$,
$\bar q_{(j)}=(\bar YY)^j \bar q$.

The authors of \cite{ILSdual} found a mapping of the baryons $B_n$ of 
(\ref{mesbaras}) to the magnetic baryons $\widetilde B_m=Y^mq^{\tilde N_c-2m}$ 
consistent with all global symmetries:
\be \label{barmapas}
    B_n \lra \widetilde B_m,\quad \w\ m=k(N_f-2)-n.
\ee

For $N_c=(2k+1)N_f-4k-1$ the magnetic theory is completely higgsed and the 
electric theory confines \cite{newconf} with low-energy spectrum given by 
the composite fields 
\bea \label{confspecas}
&&M_j,\ P_r,\  \bar P_r,\quad j=0,\ldots,k,\quad r=0,\ldots,k-1,\nonumber\\
&&B\equiv B_{k(N_f-2)},\ \bar B\equiv \bar B_{k(N_f-2)}
\eea 
of eqs.\ (\ref{mesbaras}).
Note that the baryons are of the form $B=X^{k(N_f-2)}Q^{N_f-1}$ and
therefore transform in the \afun representation of the $SU(N_f)$
flavor group. Furthermore they are mapped to the magnetic quark singlets 
that stay massless after breaking the magnetic gauge group, as can be seen
by setting $\tilde N_c=1$ in (\ref{barmapas}). The 't Hooft anomaly matching 
conditions \cite{tH} are trivially satisfied because the only fields that 
contribute to the global anomalies in the magnetic theory for $\tilde N_c=1$ 
are the $2N_f$ quark singlets and the meson singlets which carry the same 
charges as the electric baryons and mesons respectively.\footnote{On more 
general grounds it has been argued \cite{syzygies} that the matching 
conditions for the gauge invariant composite operators are satisfied 
whenever the classical constraints can be derived from a superpotential.} 
The anomaly matching conditions 
cannot tell us whether the composite fields $T_i$ of (\ref{mesbaras}) appear 
as low-energy degrees of freedom or not, because it turns out that their 
contribution to the global anomalies vanishes. The reason is that the $T_i$ 
are only charged under $U(1)_R$ and ${\bfm Z}_{2(k+1)N_f}$ and the fermionic 
charges under these symmetries have equal absolute values but opposite signs 
for each of the pairs $(T_i,T_{k+1-i})$. For the same reason each of the
bilinear terms $T_iT_{k+1-i}$ has R-charge 2 and is neutral under all
other symmetries. They should be present in the effective superpotential,
giving mass to all of the fields $T_i$.\footnote{I thank C.~Cs\'aki and 
H.~Murayama for a clarifying remark on this point.} Therefore the 
operators $T_i$ are removed from the low-energy spectrum. Another way to 
see this is to realize that the flat directions corresponding to \nonvan 
expectation values $\vev{T_i}$ (for vanishing quark VEV's) are lifted by 
the tree-level superpotential. Furthermore, the confining superpotential 
without the fields $T_i$ shown below reproduces correctly the constraints 
amongst the classical composite fields by the equations of motion (we 
checked only the constraints involving the baryons $B$, $\bar B$).

The effective low-energy superpotential of the magnetic theory, deduced
from (\ref{Wmagas}) for the theory with $\tilde N_c=2(k+1)$ and $(N_f+1)$
quark flavors by adding a tree-level term $mM_0$ and integrating out the
massive modes, contains a term $M_kq\bar q$. We thus expect that the 
confining superpotential of the electric theory has a term proportional to
\be \bar B M_k B \ \sim\ (Q\bar Q)^{N_f}(X\bar X)^{k(N_f-1)}.
\ee
Comparing this to (\ref{Weff2}) we find $\alpha=1$. From (\ref{albega})
we then get $\gamma=3-N_f$ and $\beta_X=\beta_{\bar X}=k(N_f-1)$. 
The confining superpotential consequently is of the form
\bea \label{Wconfas}
    W &= &{\bar B M_k B\ +\ {\sum_{\{r_l,j_m\},0\le p\le N_f/2}}
          \left[\prod_{l=1}^p(P_{r_l}\bar P_{r_l})
                \prod_{m=1}^{N_f-2p}M_{j_m}\right]
          \over h^{N_f-3}\,\Lambda^{4k(N_f-2)+N_f}},\\
       &&\w\ \sum_{l=1}^p(2r_l+1) + \sum_{m=1}^{N_f-2p}j_m = k(N_f-1),
       \nonumber
\eea
where we left out the relative coefficients in front of the different
terms (some of them may even vanish). They can be determined by requiring 
that the equations of motion reproduce the classical constraints.
The flavor indices of the terms containing no baryons are contracted with
two epsilon tensors of rank $N_f$. These terms are presumably generated 
by a one-instanton effect in the completely broken magnetic gauge group. 

The equations of motion for $B$, $\bar B$ of the superpotential 
(\ref{Wconfas}) give the constraints $\bar B M_k=M_k B=0$ which
are the correct relations for the classical composite operators
in the presence of the tree-level superpotential  
$\Wtree=h\,\Tr(X\bar X)^{k+1}$. To see this use an $SU(N_c)$ rotation
to transform the tensor $X$ to the normal form 
$X=diag(x_1,\ldots,x_{N_c/2})\otimes i\sigma_2$ (we assumed $N_c$ even,
the argument for $N_c$ odd is similar); let $p$ be the rank of the matrix
$X$. From the equations of motion of $\Wtree$ it then follows that the 
matrix $(\bar XX)^k$ is such that
$$((\bar XX)^k)_\alpha^{\ \beta}\bar Q^i_\beta=0\quad {\rm for}\quad 
  \beta\leq p.$$
On the other hand 
\bea &&\eps_{i_1\ldots i_{N_f}}\eps_{\alpha_1\ldots \alpha_{N_c}}
       X^{\alpha_1\alpha_2}\cdots X^{\alpha_{2k(N_f-2)-1}\alpha_{2k(N_f-2)}}
       \nonumber\\
     &&\hskip2.5cm\cdot Q^{\alpha_{2k(N_f-2)+1}i_1}\cdots 
         Q^{\alpha_{N_c}i_{N_f-1}}
         Q^{\beta i_{N_f}}=0\quad {\rm for}\quad \beta>p.\nonumber
\eea
As a consequence $M_k B$ vanishes.

There could be further terms invariant under all symmetries,
with $\alpha>1$ and possibly generated by multi-instanton effects.
However only the terms of (\ref{Wconfas}) containing the highest power of 
$1/\Lambda$ are relevant for the derivation of the classical constraints
(because the classical limit is reached when the field expectation values
are much bigger than $\Lambda$). As the classical constraints seem to follow 
from the $\alpha=1$ terms it is not likely that these multi-instanton 
corrections are present. However, we can not exclude them rigorously.

Integrating out one quark flavor we obtain a theory with $\hat N_f=N_f-1$
flavors and $\hat N_c=(2k+1)\hat N_f-2k$ colors. This theory should
possess a stable ground state because the authors of \cite{ILSdual} found
that there is a stable vacuum if $(2k+1)\hat N_f>\hat N_c$. Indeed, adding
a mass term $mM_0$ to (\ref{Wconfas}) and integrating out the
massive modes, we find a quantum moduli space with $k+1$ constraints
amongst the confined degrees of freedom $\hat M_j$, $\hat P_r$,  
$\hat{\bar P_r}$, $\hat B$, $\hat{\bar B_{\,}}$\footnote{The hats denote the 
reduction to $\hat N_f$ flavors; the baryons, e.g., are given by 
$\hat B=Q^{\hat N_f}X^{k(\hat N_f-1)}$.} of the theory with $\hat N_f$ 
quark flavors.
One of these constraints is modified quantum mechanically and reads
\bea  \label{qconstr_SU}
    \sum_{\{r_l,j_m\},\atop 0\le p\le \hat N_f/2}
         \left[\prod_{l=1}^p(\hat P_{r_l}\hat{\bar P}_{r_l})
         \prod_{m=1}^{\hat N_f-2p}\hat M_{j_m}\right]
   &= &h^{\hat N_f-2}\Lambda_L^{4k(\hat N_f-1)+\hat N_f+2},\\
  \w\ \sum_{l=1}^p(2r_l+1) + \sum_{m=1}^{\hat N_f-2p}j_m = k\hat N_f,
  \nonumber
\eea
where we again left out all relative coefficients.
This is, to our knowledge, the first example of a quantum modified moduli 
space which does not satisfy the index constraint $\Delta=0$. The argument 
of \cite{confine,qmm} to derive this constraint rested on the assumption that 
it is possible
to assign zero R-charge to all fields and then compensate the anomaly by 
assigning R-charge $\Delta$ to $\Lambda^b$. Of course this reasoning is no 
longer valid when a tree-level superpotential is present. The constraint
(\ref{qconstr_SU}) spontaneously breaks the chiral symmetry of the theory.
We checked that the 't Hooft anomaly matchings are satisfied at the point
of the moduli space where the $SU(N_f)_L\times SU(N_f)_R$ flavor symmetry
is broken to its diagonal subgroup and all $U(1)$ symmetries are unbroken.

For the special case $k=1$, $N_f=3$, and consequently $N_c=4$, this can be
seen more explicitly. For $N_c=4$ the \asym representation and its conjugate 
are equivalent. Therefore there is an additional global $SU(2)$ symmetry which
permutes the two tensors $X_1\equiv X$ and $X_2\equiv\bar X$. The mesons and 
baryons in this case are defined by $(M_0)^{ij}=Q^{ir}\bar Q^j_r$, 
$(M_1)^{ij}=\quarter\epsilon_{rstu}\epsilon^{\alpha\beta}Q^{ir}
X_\alpha^{st}X_\beta^{uv}\bar Q^j_v$, 
$B_{\alpha i}=\quarter\epsilon_{rstu}\epsilon_{ijm}Q^{jr}X_\alpha^{st}Q^{mu}$
and $\bar B_{\alpha i}=\half\epsilon_{ijm}\bar Q^j_r X_\alpha^{rs}\bar Q^m_s$,
with color indices $r,s,t,\ldots$, flavor indices $i,j,m,\ldots$ and
$SU(2)$ indices $\alpha,\beta$. The confining superpotential is
\be \label{Wconfas4}
    W={\eps^{\alpha\beta}\bar B_{\alpha i} (M_1)^{ij} B_{\beta j}\ 
     +\ \half\eps_{i_1i_2i_3}\eps_{j_1j_2j_3}
         (M_0)^{i_1j_1}(M_1)^{i_2j_2}(M_1)^{i_3j_3}\over \Lambda^7}.
\ee
Adding a mass term for the third quark flavor and integrating out the
massive modes, we find a vanishing low-energy superpotential and the
two constraints
\be \label{qconstr_SU4}
    \det \hat M_1 = \Lambda_L^8,\qquad 
    \eps^{\alpha\beta}\hat{\bar B_{\,}}_{\!\alpha} \hat B_\beta
    +\eps_{i_1i_2}\eps_{j_1j_2}(\hat M_0)^{i_1j_1}(\hat M_1)^{i_2j_2}=0
\ee
for the low-energy fields of the $\hat N_f=2$ theory.
In the limit $\Lambda_L\to 0$ these reduce to the right constraints
amongst the classical composite operators when the equations of motion
of the tree-level superpotential are taken into account.

Another interesting special case is $k=0$. The tree-level superpotential
$\Wtree=h\,\Tr X\bar X$ gives mass to the \asym tensors. The model
reduces to SQCD with $N_c=N_f-1$, which is known to confine with
a \nonpert superpotential that is smooth in the confined degrees of
freedom \cite{Seib_D49}. The superpotential (\ref{Wconfas}) has the
correct form in this limit, $W_{k=0}=(\bar B M_0 B-\det M_0)/
\Lambda_L^{2N_f-3}$, where the scale matching 
$h^{N_c-2}\Lambda^{3N_c-(N_c-2)-N_f}=\Lambda_L^{3N_c-N_f}$ was used.
The theory with one quark flavor less, i.e.\ $\hat N_c=\hat N_f$, is known 
to confine with a quantum modified moduli space \cite{Seib_D49}. The 
equation (\ref{qconstr_SU}) is modified by an additional term 
$\hat B\hat{\bar B_{\,}}$ in the limit $k=0$ and reproduces the correct 
quantum constraint $\det\hat M-\hat B\hat{\bar B}=\hat\Lambda_L^{2\hat N_f}$, 
where $h^{\hat N_f-2}\Lambda_L^{\hat N_f+2}=\hat\Lambda_L^{2N_f}$.

\subsection{$SU(N_c)$ with a symmetric tensor and its conjugate}
Consider SQCD with an additional flavor of symmetric tensors $X$, $\bar X$ 
and tree-level superpotential $\Wtree=h\,\Tr(X\bar X)^{k+1}$. 
This model was first studied in \cite{ILSdual}. The transformation
properties of the matter fields under the gauge symmetry and the \nonan
global symmetries are shown in the following table:
$$\ba{|c|c|cccccc|}
\hline    \rule[-1.3ex]{0mm}{4ex}
& SU(N_c) & SU(N_f)_L & SU(N_f)_R &U(1)_X & U(1)_B 
& U(1)_R & {\bfm Z}_{2(k+1)N_f} \\ \hline
\rule{0mm}{3ex}
Q  &\Yfun &\Yfun &\bfm 1 &0 &{1\over N_c} &1-{N_c-2k\over(k+1)N_f} &-(N_c+2)\\
\bar Q  &\Yfunb &\bfm 1 &\Yfun &0 &-{1\over N_c} 
                &1-{N_c-2k\over(k+1)N_f} &-(N_c+2)\\ 
X  &\Ysym &\bfm 1 &\bfm 1 &1 &{2\over N_c} &{1\over k+1} &N_f\\
\rule[-1.5ex]{0mm}{1.5ex}
\bar X  &\Ysymb &\bfm 1 &\bfm 1 &-1 &-{2\over N_c} &{1\over k+1} &N_f\\ \hline 
\ea$$

We will be interested in the following gauge invariant composite operators 
that can be built from the elementary fields:
\bea \label{mesbarsy}
mesons && M_j=Q\bar Q_{(j)},\quad
   P_r= Q\bar XQ_{(r)},\ \bar P_r=\bar QX\bar Q_{(r)},\nonumber\\
   &&\w Q_{(j)}=(X\bar X)^j Q,\ \bar Q_{(j)}=(\bar XX)^j\bar Q,\quad 
     j=0,\ldots,k,\quad r=0,\ldots,k-1,\nonumber \\
baryons && \cB_p^{(\bar n_0,\ldots,\bar n_{k-1},n_0,\ldots,n_k)}
        =(XW_\alpha)^2(X(\bar XX)W_\alpha)^2\cdots(X(\bar XX)^{p-1}W_\alpha)^2
        \nonumber\\
        && \hskip4.5cm \cdot(X\bar Q)^{\bar n_0}(X\bar Q_{(1)})^{\bar n_1}
           \cdots (X\bar Q_{(k-1)})^{\bar n_{k-1}}\,
           Q^{n_0}Q_{(1)}^{n_1}\cdots Q_{(k)}^{n_k},\nonumber\\
        && \bar\cB_p^{(\bar n_0,\ldots,\bar n_{k-1},n_0,\ldots,n_k)}
          =(\bar XW_\alpha)^2(\bar X(X\bar X)W_\alpha)^2\cdots
           (\bar X(X\bar X)^{p-1}W_\alpha)^2\nonumber\\
        && \hskip4.5cm \cdot(\bar XQ)^{\bar n_0}(\bar XQ_{(1)})^{\bar n_1}
           \cdots(\bar XQ_{(k-1)})^{\bar n_{k-1}}\,
           \bar Q^{n_0}\bar Q_{(1)}^{n_1}\cdots\bar Q_{(k)}^{n_k},\nonumber\\
        && \qquad\w \sum_{j=0}^kn_j+\sum_{j=0}^{k-1}\bar n_j=N_c-4p,\quad
           p=0,\ldots,\min(k,{\textstyle\left\lceil{N_c\over 4}\right\rceil}),
           \\
        && B_n=X^n Q^{N_c-n}Q^{N_c-n},\ 
           \bar B_n=\bar X^n \bar Q^{N_c-n}\bar Q^{N_c-n},\quad 
           n=0,\ldots,N_c,\nonumber\\
        && T_i=\Tr(X\bar X)^i,\quad i=1,\ldots,k,\nonumber
\eea
where the gauge indices are contracted with one epsilon tensor for the 
baryons $\cB_p^{(\cdots)}$, $\bar\cB_p^{(\cdots)}$ and with two epsilon 
tensors for the baryons $B_n$, $\bar B_n$.

This theory is dual to an $SU(\tilde N_c)$ gauge theory, with 
$\tilde N_c=(2k+1)N_f+4k-N_c$ and matter content \cite{ILSdual}
$$\ba{|c|c|cccccc|}
\hline    \rule[-1.3ex]{0mm}{4ex}
& SU(\tilde N_c) & SU(N_f)_L & SU(N_f)_R &U(1)_X & U(1)_B 
& U(1)_R & {\bfm Z}_{2(k+1)N_f} \\ \hline
\rule{0mm}{3ex}
q  &\Yfun &\Yfunb &\bfm 1 &{k(N_f+2)\over\tilde N_c} &{1\over\tilde N_c} 
          &1-{\tilde N_c-2k\over(k+1)N_f} &N_c+N_f+2\\
\bar q  &\Yfunb &\bfm 1 &\Yfunb &-{k(N_f+2)\over\tilde N_c} 
          &-{1\over\tilde N_c} &1-{\tilde N_c-2k\over(k+1)N_f} 
          &N_c+N_f+2\\ 
Y  &\Ysym &\bfm 1 &\bfm 1 &{N_c-N_f\over\tilde N_c} &{2\over\tilde N_c}
           &{1\over k+1} &N_f\\
\rule[-1.5ex]{0mm}{1.5ex}
\bar Y  &\Ysymb &\bfm 1 &\bfm 1 &-{N_c-N_f\over\tilde N_c} 
           &-{2\over\tilde N_c} &{1\over k+1} &N_f\\ \hline
\ea$$
and singlets $M_j$, $P_r$, $\bar P_r$ that carry the same quantum numbers
as the mesons of the electric theory.

The following tree-level superpotential of the magnetic theory is invariant
under all the symmetries:
\be \label{Wmagsy}
   \Wmag =-h\,\Tr (Y\bar Y)^{k+1}
          +{h\over\mu^2}\,\sum_{j=0}^k M_{k-j} q\bar q_{(j)}
          +{h\over\mu^2}\,\sum_{r=0}^{k-1} \Big[P_{k-1-r} q\bar Y q_{(r)}
                    +\bar P_{k-1-r}\bar qY\bar q_{(r)}\Big]\ ,
\ee
where $q_{(j)}=(Y\bar Y)^j q$, $\bar q_{(j)}=(\bar YY)^j \bar q$.

The electric baryons of (\ref{mesbarsy}) can be consistently mapped to 
similar baryons of the magnetic theory:
\bea \label{barmapsy}
   \cB_p^{(\bar n_0,\ldots,\bar n_{k-1},n_0,\ldots,n_k)} &\lra 
   &\widetilde\cB_q^{(\bar m_0,\ldots,\bar m_{k-1},m_0,\ldots,m_k)},
   \quad\w\nonumber\\
   &&q=k-p,\ m_j=N_f-n_{k-j},\ \bar m_j=N_f-\bar n_{k-1-j},\nonumber\\ 
   B_n &\lra &\widetilde B_m, \quad\w\ m=2k(N_f+2)-n,
\eea
where the magnetic baryons $\widetilde\cB_q^{(\ldots)}$, $\widetilde B_m$
are defined in the same way as the electric baryons of (\ref{mesbarsy}) 
replacing all fields by their dual partners and $N_c$ by $\tilde N_c$. 
The second of these mappings has been found in \cite{ILSdual}.

For $N_c=(2k+1)N_f+4k-1$ the magnetic theory is completely higgsed and the 
electric theory confines \cite{newconf} with low-energy spectrum given by 
the composite fields 
\bea \label{confspecsy}
&&M_j,\ P_r,\  \bar P_r,\quad j=0,\ldots,k,\quad r=0,\ldots,k-1,\nonumber\\
&&B\equiv \cB_k^{(N_f,\ldots,N_f,N_f-1)},\ 
\bar B\equiv \bar\cB_k^{(N_f,\ldots,N_f,N_f-1)},\\
&&b\equiv B_{2k(N_f+2)-1},\ \bar b\equiv \bar B_{2k(N_f+2)-1}\nonumber
\eea
of eqs.\ (\ref{mesbarsy}).
Note that the baryons $b$, $\bar b$ are of the form 
$b=X^{2k(N_f+2)-1}Q^{N_f}Q^{N_f}$ and therefore do not transform under the 
$SU(N_f)$ flavor group; the baryons $B$, $\bar B$ are in the \afun 
representation of $SU(N_f)$. Furthermore, from (\ref{barmapsy}) we find
the mappings $B,\bar B\lra q,\bar q$ and $b,\bar b\lra Y,\bar Y$. One color
component of each of the fields $q$, $\bar q$, $Y$, $\bar Y$ together with
the meson singlets are exactly the degrees of freedom that stay massless
after breaking the magnetic gauge group. It is easy to see that
the 't Hooft anomaly matching conditions are satisfied because the only 
fields that contribute to the global anomalies in the magnetic theory for 
$\tilde N_c=1$ are the $2N_f$ quark singlets, one component of $Y$ and 
$\bar Y$ each and the meson singlets. They carry the same charges as the 
baryons and mesons of the electric theory and consequently the global 
anomalies match between the macroscopic and the microscopic description. 
As in the model of the previous section one could add the fields $T_i$ 
without violating the matching conditions. Their contribution to the global
anomalies cancels. But in the presence of the tree-level superpotential
their VEV's do not correspond to flat directions and they are
removed from the low-energy spectrum by the mass terms $T_iT_{k+1-i}$.

The effective low-energy superpotential of the magnetic theory, deduced
from (\ref{Wmagsy}) for the theory with $\tilde N_c=2(k+1)$ and $(N_f+1)$
quark flavors by adding a tree-level term $mM_0$ and integrating out the
massive modes, contains the terms
$$\Tr(Y\bar Y)^{k+1}\ +\ M_kq\bar q\ +\ P_{k-1}q\bar Yq\ +\ 
  \bar P_{k-1}\bar qY\bar q.$$ 
We thus expect that the confining superpotential of the electric theory has 
terms proportional to
\bea \label{conftermssy}
    (b\bar b)^{k+1} &\sim 
       &(Q\bar Q)^{2(k+1)N_f}(X\bar X)^{(k+1)(2k(N_f+2)-1)},\nonumber\\
    \bar B M_k B &\sim 
       &(W_\alpha)^{4k}(Q\bar Q)^{(2k+1)N_f}(X\bar X)^{k(N_c-2k)},\\
    BBP_{k-1}\bar b,\ \bar B\bar B\bar P_{k-1}b &\sim
       &(W_\alpha)^{4k}(Q\bar Q)^{2(k+1)N_f}(X\bar X)^{(k+1)(2k(N_f+1)-1)}.
    \nonumber
\eea
Comparing this to (\ref{Weff2}) we find $\alpha^{(1)}=2(k+1)$ for the
term in the first line of (\ref{conftermssy}). From (\ref{albega}) we then
get $\gamma^{(1)}=-(N_c+N_f+4)$ and $\beta^{(1)}=(k+1)(2k(N_f+2)-1)$.
In the same way find for the terms in the second and third line
$\alpha^{(2)}=2k+1$, $\gamma^{(2)}=-N_c-2(k+1)$, $\beta^{(2)}=k(N_c-2k)$
and $\alpha^{(3)}=2(k+1)$, $\gamma^{(3)}=-(N_c+N_f+4+2k)$, 
$\beta^{(3)}=(k+1)(2k(N_f+1)-1)$, where the formula  (\ref{albega3}) has
been used with $\delta=2k$.
The confining superpotential consequently is of the form
\be \label{Wconfsy}
   W={\bar B M_k B\over h^{N_c+2(k+1)}\,\Lambda^{(2k+1)(4k(N_f+2)-4+N_f)}}
     \ +\ {(b\bar b)^{k+1}+h^{-2k}(BBP_{k-1}\bar b+\bar B\bar B\bar P_{k-1}b)
         \over h^{N_c+N_f+4}\,\Lambda^{2(k+1)(4k(N_f+2)-4+N_f)}}.
\ee
Because all mesons have negative R-charge (for $k>0$) no other superpotential 
terms with $\alpha\le2(k+1)$ are possible.
In the semiclassical regime $\Lambda\to0$ (i.e.\ only the second term of
(\ref{Wconfsy}) is relevant) the equations of motion of this superpotential 
set the expectation values of the baryons to zero. These are indeed classical 
constraints because the field strength tensor $W_\alpha$ vanishes classically.

From the analysis of \cite{ILSdual} we know that there is only a stable
vacuum if $(2k+1)N_f\ge N_c-4k$. Thus, for the theory with one less flavor
one expects to find a superpotential of the Affleck-Dine-Seiberg type which
destabilizes the ground state. It is not yet clear how this can be obtained
by integrating out one quark flavor from (\ref{Wconfsy}).

In the limit $k=0$ the coupling parameter $h$ represents a mass for the
tensors $X$, $\bar X$. Integrating out the massive modes and using the
scale matching relation $h^{N_c+2}\Lambda^{3N_c-(N_c+2)-N_f}=
\Lambda_L^{3N_c-N_f}$, we find SQCD with $N_c=N_f-1$ and the correct
confining superpotential  $W_{k=0}=(\bar B M_0 B-\det M_0)/\Lambda_L^{2N_f-3}$.
The term proportional to $\det M_0$ is only possible for $k=0$ and is
generated by a one-instanton effect in the magnetic gauge theory.
For $k=0$ there are no mesons $P_r$, $\bar P_r$ and $b$, $\bar b$ acquire 
a mass from the superpotential (\ref{Wconfsy}) as anticipated by the
authors of \cite{newconf}.

\subsection{$SU(N_c)$ with an \asym tensor and 
            a conjugate symmetric tensor}
Consider an $SU(N_c)$ gauge theory with $N_f+8$ quarks $Q$, $N_f$ 
antiquarks $\bar Q$, an \asym tensor $X$ and a conjugate symmetric
tensor $\bar X$ and tree-level superpotential 
$\Wtree=h\,\Tr(X\bar X)^{2(k+1)}$. 
This is a chiral theory and was first studied in \cite{ILSdual}.
The transformation properties of the matter fields under the gauge symmetry 
and the \nonan global symmetries are shown in the following table:
$$\ba{|c|c|cccccc|}
\hline    \rule[-1.3ex]{0mm}{4ex}
& SU(N_c) &\msmall{SU(N_f+8)_L} &\msmall{SU(N_f)_R} &U(1)_X & U(1)_B 
& U(1)_R & {\bfm Z}_{4(k+1)(N_f+4)} \\ \hline
\rule{0mm}{3ex}
Q  &\Yfun &\Yfun &\bfm 1 &\msmall{-(2k+1)+}{2(4k+3)\over N_f+8} 
          &{1\over N_c} &1-{N_c+2(4k+3)\over2(k+1)(N_f+8)} &-N_c\\
\bar Q  &\Yfunb &\bfm 1 &\Yfun &\msmall{2k+1+}{2(4k+3)\over N_f}
                &-{1\over N_c} &1-{N_c-2(4k+3)\over2(k+1)N_f} &-N_c\\ 
X  &\Yasym &\bfm 1 &\bfm 1 &1 &{2\over N_c} &{1\over2(k+1)} &N_f+4\\
\rule[-1.5ex]{0mm}{1.5ex}
\bar X  &\Ysymb &\bfm 1 &\bfm 1 &-1 &-{2\over N_c} &{1\over2(k+1)} &N_f+4\\ 
\hline 
\ea$$

We will be interested in the following gauge invariant composite operators 
that can be built from the
elementary fields:
\bea \label{mesbarsas}
mesons && M_j=Q\bar Q_{(j)},\quad
   P_r= Q\bar XQ_{(r)},\ \bar P_r=\bar QX\bar Q_{(r)},\nonumber\\
   &&\w Q_{(j)}=(X\bar X)^j Q,\ \bar Q_{(j)}=(\bar XX)^j\bar Q,\quad 
   j=0,\ldots,2k+1,\quad r=0,\ldots,2k,\nonumber \\
baryons && \bar\cB^{(\bar n_0,\ldots,\bar n_{2k},n_0,\ldots,n_{2k+1})}
          =(\bar X(X\bar X)^kW_\alpha)^2
           (\bar XQ)^{\bar n_0}(\bar XQ_{(1)})^{\bar n_1}
           \cdots(\bar XQ_{(2k)})^{\bar n_{2k}}\nonumber\\
        && \hskip4.5cm\cdot\bar Q^{n_0}\bar Q_{(1)}^{n_1}\cdots
           \bar Q_{(2k+1)}^{n_{2k+1}},\nonumber\\
        && \qquad\w \sum_{j=0}^{2k+1}n_j+\sum_{j=0}^{2k}\bar n_j=N_c-4,\\
        && B_n=X^n Q^{N_c-2n},\quad 
           n=0,\ldots,\left\lceil{N_c\over2}\right\rceil,\nonumber\\
        && \bar B_{\bar n}=\bar X^{\bar n} \bar Q^{N_c-\bar n}
               \bar Q^{N_c-\bar n},\quad \bar n=0,\ldots,N_c,\nonumber\\
        && T_i=\Tr(X\bar X)^i,\quad i=1,\ldots,2k+1,\nonumber
\eea
where the gauge indices are contracted with one epsilon tensor for the 
$\bar\cB^{(\cdots)}$, $B_n$ and with two epsilon tensors for the 
$\bar B_{\bar n}$.

This theory is dual to an $SU(\tilde N_c)$ gauge theory, with 
$\tilde N_c=(4k+3)(N_f+4)-N_c$ and matter content \cite{ILSdual}
$$\ba{|c|c|cccccc|}
\hline    \rule[-1.3ex]{0mm}{4ex}
& SU(\tilde N_c) &\msmall{SU(N_f+8)_L} &\msmall{SU(N_f)_R} &U(1)_X & U(1)_B 
& U(1)_R & {\bfm Z}_{4(k+1)(N_f+4)} \\ \hline
\rule{0mm}{3ex}
q  &\Yfun &\Yfunb &\bfm 1 &\msmall{2k+1-}{2(4k+3)\over N_f+8} 
          &{1\over\tilde N_c} &1-{\tilde N_c+2(4k+3)\over2(k+1)(N_f+8)} 
          &-\tilde N_c+p\\
\bar q  &\Yfunb &\bfm 1 &\Yfunb &\msmall{-(2k+1)-}{2(4k+3)\over N_f} 
          &-{1\over\tilde N_c} &1-{\tilde N_c-2(4k+3)\over2(k+1)N_f}
          &-\tilde N_c-p\\ 
Y  &\Yasym &\bfm 1 &\bfm 1 &-1 &{2\over\tilde N_c} &{1\over2(k+1)} &N_f+4+2p\\
\rule[-1.5ex]{0mm}{1.5ex}
\bar Y  &\Ysymb &\bfm 1 &\bfm 1 &1 &-{2\over\tilde N_c} &{1\over2(k+1)} 
                &N_f+4-2p\\ \hline 
\ea$$
and singlets $M_j$, $P_r$, $\bar P_r$ that carry the same quantum numbers
as the mesons of the electric theory. The number $p$ is defined by
$$p=4+{(2(k+1)(N_f+4)+2)(N_f+4)\over\tilde N_c}.$$

The following tree-level superpotential of the magnetic theory is invariant
under all the symmetries:
\be \label{Wmagsas}
   \Wmag =-h\,\Tr (Y\bar Y)^{2(k+1)}
          +{h\over\mu^2}\,\sum_{j=0}^{2k+1} M_{2k+1-j} q\bar q_{(j)}
          +{h\over\mu^2}\,\sum_{r=0}^{2k} \Big[P_{2k-r} q\bar Y q_{(r)}
                    +\bar P_{2k-r}\bar qY\bar q_{(r)}\Big]\ .
\ee
where $q_{(j)}=(Y\bar Y)^j q$, $\bar q_{(j)}=(\bar YY)^j \bar q$.

Under duality the electric baryons of (\ref{mesbarsas}) are mapped to 
the magnetic baryons $\widetilde{\bar\cB}{}^{(\bar m_i,m_j)}=
(\bar Yq)^{\bar m_0}\cdots(\bar Yq_{(2k)})^{\bar m_{2k}}\bar q^{m_0}\cdots
\bar q_{(2k+1)}^{m_{2k+1}}$, $\widetilde B_m=Y^mq^{\tilde N_c-2m}$ and 
$\widetilde{\bar B}_{\bar m}=Y^{\bar m}q^{\tilde N_c-\bar m}
q^{\tilde N_c-\bar m}$ according to the following prescription:
\bea \label{barmapsas}
   \bar\cB^{(\bar n_0,\ldots,\bar n_{2k},n_0,\ldots,n_{2k+1})} &\lra 
   &\widetilde{\bar\cB}{}^{(\bar m_0,\ldots,\bar m_{2k},m_0,\ldots,m_{2k+1})},
   \quad\w\nonumber\\
   &&m_j=N_f-n_{2k+1-j},\ \bar m_j=N_f+8-\bar n_{2k-j},\\ 
   B_n &\lra &\widetilde B_m,\quad \w\ m=(2k+1)(N_f+4)-2-n,\nonumber\\ 
   \bar B_{\bar n} &\lra &\widetilde{\bar B}_{\bar m},\quad \w\
   \bar m=2(2k+1)(N_f+4)+4-\bar n.\nonumber
\eea
The last two of these mappings have been found in \cite{ILSdual}.

For $N_c=(4k+3)(N_f+4)-1$ the magnetic theory is completely higgsed and the 
electric theory confines with low-energy spectrum given by the composite 
fields 
\bea \label{confspecsas} 
&&M_j,\ P_r,\  \bar P_r,\quad j=0,\ldots,2k+1,\quad r=0,\ldots,2k,\nonumber\\
&&B\equiv B_{(2k+1)(N_f+4)-2},\ 
\bar B\equiv\cB^{(N_f+8,\ldots,N_f+8,N_f,\ldots,N_f,N_f-1)},\\ 
&&\bar b\equiv \bar B_{2(2k+1)(N_f+4)+3},\nonumber
\eea
of eqs.\ (\ref{mesbarsas}).

The baryons $B$, $\bar B$ transform in the \afun representation of 
$SU(N_f+8)_L$, $SU(N_f)_R$ respectively and $\bar b$ does not transform
under the flavor symmetry. The fact that the baryons are
very different from the antibaryons is due to the chirality of the theory.
The former resemble the baryons of the theory with an \asym flavor whereas
the latter are similar to the antibaryons of the theory with a symmetric
flavor. From (\ref{barmapsas}) we find the mappings $B,\bar B\lra q,\bar q$
and $\bar b\lra \bar Y$. One color component of each of the fields $q$,
$\bar q$, $\bar Y$ together with the meson singlets are exactly the degrees 
of freedom that stay massless
after breaking the magnetic gauge group. The 't Hooft anomaly matching 
conditions are satisfied because the only fields that contribute to the 
global anomalies in the magnetic theory for $\tilde N_c=1$ are the $2N_f$ 
quark singlets, one component of $\bar Y$  and the meson singlets. 
The contribution of the $T_i$ to the global anomalies again vanishes 
because they are only charged under $U(1)_R$ and ${\bfm Z}_{4(k+1)(N_f+4)}$ 
and the fermionic charges under these symmetries have equal absolute values 
but opposite signs for each of the pairs $(T_i,T_{2(k+1)-i})$. 
However, in the presence of the tree-level superpotential their VEV's do 
not correspond to flat directions and they are removed from the 
low-energy spectrum by the mass terms $T_iT_{2(k+1)-i}$.

As a further consistency check let us consider deformations of the theory
along the flat directions corresponding to large expectation values of the 
baryons $B$, $\bar b$. A large VEV of $B$ breaks
the gauge symmetry to $Sp(2((2k+1)(N_f+4)-2))$ \cite{ILSdual}. The low-energy
theory contains $2(N_f+4)$ quarks $Q$, a symmetric tensor $X$ and tree-level
superpotential $\Tr X^{2(k+1)}$. This model is known to show confinement
\cite{newconf} (cf.\ section 3.4). A large VEV of $\bar b$ breaks the
gauge symmetry to $SO(2(2k+1)(N_f+4)+3)$ \cite{ILSdual}. The low-energy
theory contains $2(N_f+4)$ quarks $Q$, an \asym tensor $X$ and tree-level
superpotential $\Tr X^{2(k+1)}$. This model is known to show confinement
\cite{newconf} (cf.\ section 3.6).

The effective low-energy superpotential of the magnetic theory, deduced
from (\ref{Wmagsas}) for the theory with $\tilde N_c=4(k+1)$, $(N_f+9)$ quarks
and $(N_f+1)$ antiquarks by adding a tree-level term $mM_0$ and integrating 
out the massive modes, contains the terms 
$M_{2k+1}q\bar q\ +\ P_{2k}q\bar Yq$. 
We thus expect that the confining superpotential of the electric theory has 
terms proportional to
\bea \label{conftermssas}
    \bar B M_{2k+1} B &\sim &(W_\alpha)^2Q^{2(k+1)(N_f+8)}\bar Q^{2(k+1)N_f}
    X^{2(k+1)((2k+1)(N_f+4)-3)}\bar X^{2(k+1)((2k+1)(N_f+4)+1)},\nonumber\\
    BBP_{2k}\bar b &\sim &Q^{2(N_f+8)}\bar Q^{2N_f}
    X^{2(2k+1)(N_f+4)-4+2k}\bar X^{2(2k+1)(N_f+4)+4+2k}.
\eea
Comparing this to (\ref{Weff2}) we find $\alpha^{(1)}=2(k+1)$ for the term
in the first line of (\ref{conftermssas}). From (\ref{albega}, \ref{albega3}) 
we then get $\gamma^{(1)}=-2(k+1)(N_f+4)$ and 
$\beta^{(1)}_X=2(k+1)((2k+1)(N_f+4)-3)$, 
$\beta^{(1)}_{\bar X}=2(k+1)((2k+1)(N_f+4)+1)$. In the same way we obtain
$\alpha^{(2)}=2$, $\gamma^{(2)}=1-2(N_f+4)$,
$\beta^{(2)}_X=2(2k+1)(N_f+4)-4+2k$, 
$\beta^{(2)}_{\bar X}=2(2k+1)(N_f+4)+4+2k$  for the term in the second line.
The confining superpotential consequently is of the form
\be \label{Wconfsas}
    W={\bar B M_{2k+1} B\over h^{2(k+1)(N_f+4)}\,
       \Lambda^{2(k+1)((8k+5)(N_f+4)-2)}}\ +\ 
      {BBP_{2k}\bar b\over h^{2(N_f+4)-1}\Lambda^{2((8k+5)(N_f+4)-2)}}\ +\ 
      \ldots,
\ee
where the dots stand for possible further terms that could be generated by
instanton effects in the completely broken magnetic gauge group.

\subsection{$Sp(2N_c)$ with an adjoint tensor}
Consider an $Sp(2N_c)$ gauge theory with $2N_f$ quarks $Q$ in the fundamental
representation and a second rank tensor $X$ in the symmetric (=adjoint)
representation of the gauge group and tree-level superpotential 
$\Wtree=h\,\Tr X^{2(k+1)}$. This model was first studied in \cite{LS}. 
The transformation properties of the matter fields under the gauge symmetry 
and the \nonan global symmetries are shown in the following table:
$$\ba{|c|c|ccc|}
\hline    \rule[-1.3ex]{0mm}{4ex}
& Sp(2N_c) & SU(2N_f) & U(1)_R & {\bfm Z}_{2(k+1)N_f} \\ \hline
\rule{0mm}{3ex}
Q  &\Yfun &\Yfun &1-{N_c+1\over(k+1)N_f} &-(N_c+1)\\
\rule[-1.5ex]{0mm}{1.5ex}
X  &\Ysym &\bfm 1 &{1\over k+1} &N_f\\ \hline
\ea$$

There are no baryons in symplectic gauge theories and therefore the only
(non-redundant) gauge invariant composite operators that can be built from 
the elementary fields are:
\bea \label{messpsy}
       && M_j=QX^jQ,\quad j=0,\ldots,2k,\nonumber\\
       && T_i=\Tr X^{2i},\quad i=1,\ldots,k,
\eea
where the gauge indices are contracted with the $Sp(2N_c)$-invariant
$J$-tensor.

This theory is dual to an $Sp(2\tilde N_c)$ gauge theory, with 
$\tilde N_c=(2k+1)N_f-2-N_c$ and matter content \cite{LS}
$$\ba{|c|c|ccc|}
\hline    \rule[-1.3ex]{0mm}{4ex}
& Sp(2\tilde N_c) &SU(2N_f) & U(1)_R & {\bfm Z}_{2(k+1)N_f} \\ \hline
\rule{0mm}{3ex}
q  &\Yfun &\Yfunb &1-{\tilde N_c+1\over(k+1)N_f} &N_c+N_f+1\\
\rule[-1.5ex]{0mm}{1.5ex}
Y  &\Ysym &\bfm 1 &{1\over k+1} &N_f\\ \hline
\ea$$
and singlets $M_j$ that carry the same quantum numbers as the mesons of the 
electric theory.

The following tree-level superpotential of the magnetic theory is invariant
under all the symmetries:
\be \label{Wmagspsy}
   \Wmag =-h\,\Tr Y^{2(k+1)}
          +{h\over\mu^2}\,\sum_{j=0}^{2k} M_{2k-j} qY^jq.
\ee

For $N_c=(2k+1)N_f-2$ the magnetic theory is completely higgsed and the 
electric theory confines \cite{newconf} with low-energy spectrum given by 
the composite fields 
\be \label{confspecspsy} M_j,\quad j=0,\ldots,2k, 
\ee
of eqs.\ (\ref{messpsy}).
The 't Hooft anomaly matching conditions are trivially satisfied because
the only fields that stay massless after completely breaking the magnetic
gauge group are the meson singlets which carry the same charges as the
mesons of the electric theory. Again the fields $T_i$ could be added to
the confined spectrum without modifying the anomaly matchings. But for the
same reason as in the models considered in the previous sections
they are not the right degrees of freedom of the low-energy theory.

The fact that all components of the magnetic quarks get massive when
completely breaking the magnetic gauge group makes it more difficult to
find the confining superpotential because none of the magnetic tree-level
terms survives the symmetry breaking. All the terms of the confining 
superpotential are generated by instanton effects. However, for the model
considered in this section we can use the method developed in \cite{newconf}
to determine the constraints on the confined degrees of freedom and then
try to find a superpotential that reproduces these constraints via the
equations of motion. To find the classical constraints we introduce 
dressed quarks $Q_{(j)}=X^jQ$ and view the considered model as an
$Sp(2((2k+1)N_f-2))$ gauge theory with $2(2k+1)N_f$ quarks 
$\cQ=(Q,Q_{(1)},\ldots,Q_{(2k)})$ and no tensor. The classical constraints 
for this reduced theory are known \cite{IP}: The mesons $\cM=\cQ\cQ$
must verify $\eps_{i_1\cdots i_{2(2k+1)N_f}}\cM^{i_3i_4}\cdots
\cM^{i_{2(2k+1)N_f-1}i_{2(2k+1)N_f}}=0$, $i_1,i_2=1,\ldots,2(2k+1)N_f$.
In \cite{newconf} the constraints for the special case $k=1$, $N_f=1$
were explicitly determined in terms of the $M_j$ and the superpotential that
reproduces these constraints was constructed. In general the terms of 
the confining superpotential should consist of products of $(2k+1)N_f$ mesons
$\cM$. Then one can expect that the equations $\partial W/\partial\cM^{ij}=0$ 
give the classical constraints. Comparing this to (\ref{Weff2}) we obtain
$\alpha=2k+1$ and from  (\ref{albega}) $\gamma=-(N_c+1)$, $\beta=2k(N_c+1)$.
 
The confining superpotential consequently is of the form
\bea \label{Wconfspsy}
    W &= &{\sum_{\{j_l\}}\prod_{l=1}^{(2k+1)N_f}M_{j_l}\over h^{N_c+1}\,
       \Lambda^{(2k+1)((4k+1)N_f-2)}},\\
    \w && \sum_{l=1}^{(2k+1)N_f}\!\!j_l=2k((2k+1)N_f-1),\nonumber
\eea
where the flavor indices are contracted with $(2k+1)$ epsilon tensors
of rank $2N_f$.

From the analysis of \cite{LS} we know that this model has a stable vacuum
if $(2k+1)N_f\ge N_c+1$. Thus, the theory with $2\hat N_f=2(N_f-1)$ quarks
and $\hat N_c=(2k+1)\hat N_f+2k-1$, which is obtained by integrating out
two quarks, does not possess a stable ground state.

In the limit $k=0$ the tree-level term $h\,\Tr X^{2(k+1)}$ gives mass to
the adjoint tensor. Integrating it out we find an $Sp(2N_c)$ gauge theory
with $N_c=N_f-2$ and the correct \cite{IP} confining superpotential
$W_{k=0}=\Pf M_0/\Lambda_L^{2N_f-3}$, where we used the scale matching
relation $h^{N_c+1}\Lambda^{2(N_c+1)-N_f}=\Lambda_L^{3(N_c+1)-N_f}$.

\subsection{$Sp(2N_c)$ with an \asym tensor}
Consider an $Sp(2N_c)$ gauge theory with $2N_f$ quarks $Q$ in the fundamental
representation and a second rank tensor $X$ in the traceless (i.e.\ 
$J_{ab}X^{ba}=0$) \asym representation of the
gauge group and tree-level superpotential $\Wtree=h\,\Tr X^{k+1}$. 
This model was first studied in \cite{Intri}. 
The transformation properties of the matter fields under the gauge symmetry 
and the \nonan global symmetries are shown in the following table:
$$\ba{|c|c|ccc|}
\hline    \rule[-1.3ex]{0mm}{4ex}
& Sp(2N_c) & SU(2N_f) & U(1)_R & {\bfm Z}_{(k+1)N_f} \\ \hline
\rule{0mm}{3ex}
Q  &\Yfun &\Yfun &1-{2(N_c+k)\over(k+1)N_f} &-(N_c-1)\\
\rule[-1.5ex]{0mm}{1.5ex}
X  &\Yasym &\bfm 1 &{2\over k+1} &N_f\\ \hline
\ea$$

The (non-redundant) gauge invariant composite operators that can be built from
the elementary fields are:
\bea \label{messpas}
       && M_j=QX^jQ,\quad j=0,\ldots,k-1,\nonumber\\
       && T_i=\Tr X^i,\quad i=2,\ldots,k,
\eea
where the gauge indices are contracted with the $Sp(2N_c)$-invariant
$J$-tensor.

This theory is dual to an $Sp(2\tilde N_c)$ gauge theory, with 
$\tilde N_c=k(N_f-2)-N_c$ and matter content \cite{Intri}
$$\ba{|c|c|ccc|}
\hline    \rule[-1.3ex]{0mm}{4ex}
& Sp(2\tilde N_c) &SU(2N_f) & U(1)_R & {\bfm Z}_{(k+1)N_f} \\ \hline
\rule{0mm}{3ex}
q  &\Yfun &\Yfunb &1-{2(\tilde N_c+k)\over(k+1)N_f} &N_c+N_f-1\\
\rule[-1.5ex]{0mm}{1.5ex}
Y  &\Yasym &\bfm 1 &{2\over k+1} &N_f\\ \hline
\ea$$
and singlets $M_j$ that carry the same quantum numbers as the mesons of the 
electric theory.

The following tree-level superpotential of the magnetic theory is invariant
under all the symmetries:
\be \label{Wmagspas}
   \Wmag =-h\,\Tr Y^{k+1}+{h\over\mu^2}\,\sum_{j=0}^{k-1} M_{k-1-j} qY^jq.
\ee

For $N_c=k(N_f-2)$ the magnetic theory is completely higgsed and the 
electric theory confines \cite{newconf} with low-energy spectrum given by 
the composite fields 
\be \label{confspecspas} M_j,\quad j=0,\ldots,k-1,\quad T_k
\ee 
of eqs.\ (\ref{messpas}). It is easy to see that the 't Hooft anomaly matching
conditions are satisfied. The contribution of $Y$ to the global anomalies is
$(\tilde N_c(2\tilde N_c-1)-1)$ times its fermionic charge under the 
considered symmetry. For $\tilde N_c=0$ it acts therefore like one field with
the negative of the charge of $Y$. Thus, we have to search for a composite
field of the electric theory that carries fermionic charge of the same
absolute value as $Y$ but of the opposite sign . This condition is satisfied by
$T_k$. The only other contribution to the global anomalies in the magnetic
theory for $\tilde N_c=0$ comes from the meson singlets which carry the
same charges as the mesons of the electric theory. The fields $T_i$,
$i=2,\ldots,k-1$, do not correspond to classical flat directions. They
are removed from the low-energy spectrum by mass terms $T_iT_{k+1-i}$.

Like for the model of the previous section none of the magnetic tree-level
terms survives the symmetry breaking. In addition, the method of considering 
the dressed quarks as the only degrees of freedom to determine the classical
constraints does not work for $Sp(2N_c)$ with an \asym tensor.
For the theory without tree-level superpotential these constraints
have been found in \cite{CSS} for $N_f=3$ and small values of $N_c$.
As the effect of the \nonvan tree-level superpotential is to remove the
$T_2,\ldots,T_{k-1}$ from the moduli space, we expect to find the
classical constraints with $\Wtree\neq0$ by setting to zero all the terms in
the classical constraints of the model with $\Wtree=0$. All the confining
superpotentials determined in \cite{CSS} are generated by one-instanton
effects. Therefore, it is likely that the confining superpotential of
the theory with $\Wtree\neq0$ is also due to a one-instanton effect.
Furthermore, one can flow from the $SU(N_c)$ gauge theory with an
\asym flavor to the model considered in this section by giving a large
expectation value to the baryon $B$ of (\ref{confspecas}). The quarks 
should thus be raised to the same power in the confining superpotentials
of both models. For the power $\alpha$ appearing in (\ref{Weff2}) this means 
$\alpha=1$. From (\ref{albega}) we find $\gamma=3-N_f$ and 
$\beta=(k-1)(N_f-1)$. This leads us to the following \nonpert superpotential 
which is invariant under all symmetries:
\bea  \label{Wconfspas}
    W &= &{\sum_{\{j_l\},p}(T_k)^p\prod_{l=1}^{N_f}M_{j_l}\over h^{N_f-3}\,
       \Lambda^{(2k-1)N_f-4(k-1)}},\\
    \w && \sum_{l=1}^{N_f}j_l=(k-1)(N_f-1)-pk,\nonumber
\eea
where the flavor indices are contracted with an epsilon tensor.

Adding a mass term $mM_0$ for two quarks to the theory with $N_c=k(N_f-2)$ 
and integrating out the massive modes we get a low-energy theory with 
$2\hat N_f=2(N_f-1)$ quarks and $\hat N_c=k(\hat N_f-1)$. From the analysis 
of \cite{Intri} we know that this model has a stable vacuum if 
$k\hat N_f>\hat N_c$. This condition is clearly satisfied. Indeed, we
find a quantum moduli space with $k$ constraints amongst the confined degrees 
of freedom $\hat M_j$, $\hat T_k$ of the theory with $2\hat N_f$ quarks.
One of these constraints is modified quantum mechanically and reads
\bea  \label{qconstr_Sp}
    \sum_{\{j_l\},p}(T_k)^p\prod_{l=1}^{\hat N_f}\hat M_{j_l}
   &= &h^{\hat N_f-2}\Lambda_L^{(2k-1)\hat N_f-2(k-2)},\\
  \w\ &&\sum_{l=1}^{\hat N_f}j_l =(k-1)\hat N_f-pk.\nonumber
\eea
The constraint (\ref{qconstr_Sp}) spontaneously breaks the chiral symmetry 
of the theory. We checked that the 't Hooft anomaly matchings are satisfied 
at the point of the moduli space where the $SU(2N_f)$ flavor symmetry
is broken to its $Sp(2N_f)$ subgroup.

In the limit $k=1$ the tree-level term $h\,\Tr X^{k+1}$ gives mass to
the \asym tensor. Integrating it out we find an $Sp(2N_c)$ gauge theory
with $N_c=N_f-2$ and the correct \cite{IP} confining superpotential
$W_{k=1}=\Pf M_0/\Lambda_L^{2N_f-3}$, where we used the scale matching
relation $h^{N_c-1}\Lambda^{3(N_c+1)-(N_c-1)-N_f}=\Lambda_L^{3(N_c+1)-N_f}$.
The theory with two quarks less, i.e.\ $\hat N_c=\hat N_f-1$, is known to
confine with a quantum modified moduli space \cite{IP}. The equation
(\ref{qconstr_Sp}) reproduces the correct quantum constraint
$\Pf\hat M=\hat\Lambda_L^{2\hat N_f}$, where 
$h^{\hat N_f-2}\Lambda_L^{\hat N_f+2}=\hat\Lambda_L^{2N_f}$.

\subsection{$SO(N_c)$ with an adjoint tensor}
Consider an $SO(N_c)$ gauge theory with $N_f$ quarks $Q$ in the fundamental
representation and a second rank tensor $X$ in the \asym (=adjoint) 
representation of the gauge group and tree-level superpotential 
$\Wtree=h\,\Tr X^{2(k+1)}$. This model was first studied in \cite{LS}. 
The transformation properties of the matter fields under the gauge symmetry 
and the \nonan global symmetries are shown in the following table:
$$\ba{|c|c|cccc|}
\hline    \rule[-1.3ex]{0mm}{4ex}
& SO(N_c) & SU(N_f) & U(1)_R & {\bfm Z}_{2(k+1)N_f} & {\bfm Z}_{2N_f} \\ \hline
\rule{0mm}{3ex}
Q  &\Yfun &\Yfun &1-{N_c-2\over(k+1)N_f} &-(N_c-2) &1\\
\rule[-1.5ex]{0mm}{1.5ex}
X  &\Yasym &\bfm 1 &{1\over k+1} &N_f &0\\ \hline
\ea$$

We will be interested in the following gauge invariant composite operators 
that can be built from the
elementary fields:
\bea \label{mesbarsoas}
mesons   && M_j=QX^jQ,\quad j=0,\ldots,2k,\nonumber\\
baryons  && B_n=X^nQ^{N_c-2n},\quad 
            n=0,\ldots,\left\lceil{N_c\over2}\right\rceil,\\
         && T_i=\Tr X^{2i},\quad i=1,\ldots,k,\nonumber
\eea
where the gauge indices are contracted with a Kronecker delta for the 
mesons and $T_i$ and with an epsilon tensor for the baryons.

This theory is dual to an $SO(\tilde N_c)$ gauge theory, with 
$\tilde N_c=(2k+1)N_f+4-N_c$ and matter content \cite{LS}
$$\ba{|c|c|cccc|}
\hline    \rule[-1.3ex]{0mm}{4ex}
& SO(\tilde N_c) &SU(N_f) & U(1)_R & {\bfm Z}_{2(k+1)N_f} & {\bfm Z}_{2N_f} \\
\hline\rule{0mm}{3ex}
q  &\Yfun &\Yfunb &1-{\tilde N_c-2\over(k+1)N_f} &N_c+N_f-2 &-1\\
\rule[-1.5ex]{0mm}{1.5ex}
Y  &\Yasym &\bfm 1 &{1\over k+1} &N_f &0\\ \hline
\ea$$
and singlets $M_j$ that carry the same quantum numbers as the mesons of the 
electric theory.

The following tree-level superpotential of the magnetic theory is invariant
under all the symmetries:
\be \label{Wmagsoas}
   \Wmag =-h\,\Tr Y^{2(k+1)}+{h\over\mu^2}\,\sum_{j=0}^{2k} M_{2k-j} qY^jq.
\ee

Under duality the electric baryons of (\ref{mesbarsoas}) are mapped to 
the magnetic baryons $\widetilde B_m=Y^mq^{\tilde N_c-2m}$ 
according to
\be \label{barmapsoas}
   B_n\ \lra\  \widetilde B_m,\quad \w m=kN_f+2-n. 
\ee

For $N_c=(2k+1)N_f+3$ the magnetic theory is completely higgsed and the 
electric theory confines \cite{newconf} with low-energy spectrum given by 
the composite fields 
\bea \label{confspecsoas}  
     &&M_j,\quad j=0,\ldots,2k,\nonumber\\
     &&B\equiv B_{kN_f+2} 
\eea
of eqs.\ (\ref{mesbarsoas}). The baryons $B$ are of the form 
$B=X^{kN_f+2}Q^{N_f-1}$ and transform in the \afun representation of 
$SU(N_f)$. Furthermore from (\ref{barmapsoas}) we find that they are mapped
to the $N_f$ magnetic quark singlets $q$ that stay massless after breaking 
the magnetic gauge group. It is easy to see that the 't Hooft anomaly 
matching conditions are satisfied because the only fields that contribute 
to the global anomalies in the magnetic theory for $\tilde N_c=1$ are the 
$N_f$ quark singlets and the meson singlets. They carry the same charges as 
the baryons and mesons of the electric theory and consequently the global 
anomalies match between the macroscopic and the microscopic description. 
As in the models of the previous sections one could add the fields $T_i$ 
without violating the matching conditions. But they are removed from the 
low-energy spectrum by the mass terms $T_iT_{k+1-i}$.

The effective low-energy superpotential of the magnetic theory, deduced
from (\ref{Wmagsoas}) for the theory with $\tilde N_c=2(k+1)$ and $(N_f+1)$
quarks by adding a tree-level term $mM_0$ and integrating out the
massive modes, contains a term $M_{2k}qq$. We thus expect that the 
confining superpotential of the electric theory has a term proportional to
\be B M_{2k} B\ \sim\ Q^{2N_f}X^{2k(N_f+1)+4}.
\ee
Comparing this to (\ref{Weff2}) we find $\alpha=1$. From (\ref{albega}) 
we then get $\gamma=1-N_f$ and $\beta=2k(N_f+1)+4$. 
The confining superpotential consequently is of the form
\be \label{Wconfsoas}
    W={B M_{2k} B\over h^{N_f-1}\,
      \Lambda^{(4k+1)N_f+2}}.
\ee
Instanton corrections may modify this superpotential by multiplying it with
an holomorphic function of the expression
\be \label{Wsoasmodif}
{\left(\prod_{l=1}^{N_f}M_{j_l}\right)BM_{2k}B\over 
      h^{2N_f}\Lambda^{2(4k+1)N_f+4}},\qquad\wh\ 
      \sum_{l=1}^{N_f}j_l=2k(N_f-1).
\ee
The equations of motion of this superpotential give $M_{2k}B=0$. 
This is indeed a classical constraint because from the equations
of motion of the tree-level superpotential it follows that $X^{2k+1}=0$.
Thus, $X$ can at most have rank $2k$. As a consequence a totally 
antisymmetrized product of more than $2k$ factors of matrix elements
of $X$ must vanish. This means that $\vev B$=0 classically.
(For $N_f=1$ it vanishes because in this case $B=\Pf X=\sqrt{\det X}=0$.)

From the analysis of \cite{LS} we know that there is only a stable
vacuum if $(2k+1)N_f\ge N_c-4$. Thus, for the theory with $\hat N_f=N_f-1$
quarks and $\hat N_c=(2k+1)\hat N_f+2k+4$ one expects to find a 
superpotential of the Affleck-Dine-Seiberg type which destabilizes the 
ground state. We did not check this but suppose that it can be derived
by integrating out one quark from (\ref{Wconfsoas}) when the corrections
(\ref{Wsoasmodif}) are taken into account. 

In the limit $k=0$ the coupling parameter $h$ represents a mass for the
tensor $X$. Integrating it out and using the scale matching relation 
$h^{N_c-2}\Lambda^{2(N_c-2)-N_f}=\Lambda_L^{3(N_c-2)-N_f}$, we find 
an $SO(N_c)$ gauge theory with $N_c=N_f+3$ and the correct \cite{SOdual}
confining superpotential after replacing the operator $X^2$ in $B$ by
$(W_\alpha)^2/h$, i.e.\ setting $B^\prime\equiv(W_\alpha)^2Q^{N_f-1}=hB$,
as explained in \cite{newconf},
$W_{k=0}=B^\prime M_0 B^\prime/\Lambda_L^{2N_f+3}$.

\subsection{$SO(N_c)$ with a symmetric tensor}
Consider an $SO(N_c)$ gauge theory with $N_f$ quarks $Q$ in the fundamental
representation and a second rank tensor $X$ in the traceless symmetric 
representation of the gauge group and tree-level superpotential 
$\Wtree=h\,\Tr X^{k+1}$. This model was first studied in \cite{Intri}. 
The transformation properties of the matter fields under the gauge symmetry 
and the \nonan global symmetries are shown in the following table:
$$\ba{|c|c|cccc|}
\hline    \rule[-1.3ex]{0mm}{4ex}
& SO(N_c) & SU(N_f) & U(1)_R & {\bfm Z}_{(k+1)N_f} & {\bfm Z}_{2N_f} \\ \hline
\rule{0mm}{3ex}
Q  &\Yfun &\Yfun &1-{2(N_c-2k)\over(k+1)N_f} &-(N_c+2) &1\\
\rule[-1.5ex]{0mm}{1.5ex}
X  &\Ysym &\bfm 1 &{2\over k+1} &N_f &0\\ \hline
\ea$$

We will be interested in the following gauge invariant composite operators 
that can be built from the
elementary fields:
\bea \label{mesbarsosy}
mesons   && M_j=QQ_{(j)},\quad\w\ Q_{(j)}=X^jQ,\quad j=0,\ldots,k-1,\nonumber\\
baryons  && \cB_p^{(n_0,\ldots,n_{k-1})}
          =(W_\alpha)^2(XW_\alpha)^2\cdots(X^{p-1}W_\alpha)^2
            Q^{n_0}Q_{(1)}^{n_1}\cdots Q_{(k-1)}^{n_{k-1}},\nonumber\\
        && \qquad\w \sum_{j=0}^{k-1}n_j=N_c-4p,\quad
           p=0,\ldots,\min(k,\textstyle\left\lceil{N_c\over 4}\right\rceil),\\
        && T_i=\Tr X^i,\quad i=2,\ldots,k-1,\nonumber
\eea
where the gauge indices are contracted with a Kronecker delta for the mesons
and $T_i$ and with an epsilon tensor for the baryons.

This theory is dual to an $SO(\tilde N_c)$ gauge theory, with 
$\tilde N_c=k(N_f+4)-N_c$ and matter content \cite{Intri}
$$\ba{|c|c|cccc|}
\hline    \rule[-1.3ex]{0mm}{4ex}
& SO(\tilde N_c) &SU(N_f) & U(1)_R & {\bfm Z}_{(k+1)N_f} & {\bfm Z}_{2N_f} \\ 
\hline\rule{0mm}{3ex}
q  &\Yfun &\Yfunb &1-{2(\tilde N_c-2k)\over(k+1)N_f} &N_c+N_f+2 &-1\\
\rule[-1.5ex]{0mm}{1.5ex}
Y  &\Ysym &\bfm 1 &{2\over k+1} &N_f &0\\ \hline
\ea$$
and singlets $M_j$ that carry the same quantum numbers as the mesons of the 
electric theory.

The following tree-level superpotential of the magnetic theory is invariant
under all the symmetries:
\be \label{Wmagsosy}
   \Wmag =-h\,\Tr Y^{k+1}+{h\over\mu^2}\,\sum_{j=0}^{k-1} M_{k-1-j} qY^jq.
\ee

The electric baryons of (\ref{mesbarsosy}) can be consistently mapped to 
similar baryons of the magnetic theory:
\be \label{barmapsosy}
   \cB_p^{(n_0,\ldots,n_{k-1})}\ \lra\ \widetilde\cB_q^{(m_0,\ldots,m_{k-1})},
   \quad\w\ q=k-p,\ m_j=N_f-n_{k-1-j}, 
\ee
where the magnetic baryons $\widetilde\cB_q^{(\ldots)}$
are defined in the same way as the electric baryons of (\ref{mesbarsosy}) 
replacing all fields by their dual partners and $N_c$ by $\tilde N_c$. 
(This mapping is similar to the one that has been found in \cite{Intri}.)

For $N_c=k(N_f+4)-1$ the magnetic theory is completely higgsed and the 
electric theory confines \cite{newconf} with low-energy spectrum given by 
the composite fields 
\bea \label{confspecsosy}  
      &&M_j,\quad j=0,\ldots,k-1,\nonumber\\
      &&B\equiv\cB_k^{(N_f,\ldots,N_f,N_f-1)}
\eea
of eqs.\ (\ref{mesbarsosy}).  The baryons $B$ transform in the \afun 
representation of $SU(N_f)$. From (\ref{barmapsosy}) we find that 
they are mapped to the $N_f$ magnetic quark singlets $q$ that stay 
massless after breaking the magnetic gauge group. It is easy to see that 
the 't Hooft anomaly matching conditions are satisfied because the only 
fields that contribute to the global anomalies in the magnetic theory for 
$\tilde N_c=1$ are the $N_f$ quark singlets and the meson singlets which
carry the same charges as the baryons and mesons of the electric theory.
As in the models of the previous sections one could add the fields $T_i$ 
without violating the matching conditions. But they are removed from the 
low-energy spectrum by the mass terms $T_iT_{k+1-i}$.

The effective low-energy superpotential of the magnetic theory, deduced
from (\ref{Wmagsosy}) for the theory with $\tilde N_c=k+1$ and $(N_f+1)$
quarks by adding a tree-level term $mM_0$ and integrating out the
massive modes, contains a term $M_{k-1}qq$. We thus expect that the 
confining superpotential of the electric theory has a term proportional to
\be B M_{k-1} B\ \sim\ (W_\alpha)^{4k}Q^{2kN_f}X^{(k-1)(N_c-2k)}.
\ee
Comparing this to (\ref{Weff2}) we find $\alpha=k$. From (\ref{albega},
\ref{albega3}) we then get $\gamma=-(N_c+2k)$ and $\beta=(k-1)(N_c-2k)$. 
The confining superpotential consequently is of the form
\be \label{Wconfsosy}
    W={B M_{k-1} B\over h^{N_c+2k}\,
      \Lambda^{k((2k-1)N_f+8k-10))}}.
\ee
The equations of motion of this superpotential give $M_{k-1}B=0$. 
This is the correct result in the classical limit because
the field strength tensor $W_\alpha$ vanishes classically. 

From the analysis of \cite{Intri} we know that there is only a stable
vacuum if $kN_f\ge N_c-4k$. Thus, for the theory with $\hat N_f=N_f-1$
quarks and $\hat N_c=k(\hat N_f+5)-1$ one expects to find a 
superpotential of the Affleck-Dine-Seiberg type which destabilizes 
the ground state. Possibly this can be obtained by integrating out 
one quark from (\ref{Wconfsosy}) when instanton corrections are taken
into account.

In the limit $k=1$ the coupling parameter $h$ represents a mass for the
tensor $X$. Integrating it out and using the scale matching relation 
$h^{N_c+2}\Lambda^{3(N_c-2)-(N_c+2)-N_f}=\Lambda_L^{3(N_c-2)-N_f}$, we find 
an $SO(N_c)$ gauge theory with $N_c=N_f+3$ and the correct \cite{SOdual}
confining superpotential $W_{k=1}=B M_0 B/\Lambda_L^{2N_f+3}$.

\section{Models with $D_{k+2}$-type superpotentials}
\subsection{$SU(N_c)$ with two adjoint tensors}
Consider SQCD with two additional second rank tensors $X$, $Y$, both
transforming in the adjoint representation, and tree-level superpotential
$\Wtree=h_1\,\Tr X^{k+1}\ +\ h_2\,\Tr XY^2$. This model was first studied 
in \cite{Bro}. The transformation properties of the matter fields under 
the gauge symmetry and the \nonan global symmetries are shown in the 
following table:
$$\ba{|c|c|cccccc|}
\hline  \rule[-1.3ex]{0mm}{4ex}
& SU(N_c) & SU(N_f)_L & SU(N_f)_R & U(1)_B 
& U(1)_R & {\bfm Z}_{2(k+1)N_f} & {\bfm Z}_{2N_f} \\ \hline
\rule{0mm}{3ex}
Q  &\Yfun &\Yfun &\bfm 1 &{1\over N_c} &1-{N_c\over(k+1)N_f} &-N_c &-N_c\\
\bar Q  &\Yfunb &\bfm 1  &\Yfun &-{1\over N_c} &1-{N_c\over(k+1)N_f} 
                &-N_c &-N_c\\ 
X  &adj &\bfm 1 &\bfm 1 &0 &{2\over k+1} &2N_f &0\\
\rule[-1.5ex]{0mm}{1.5ex}
Y  &adj &\bfm 1 &\bfm 1 &0 &{k\over k+1} &-N_f &N_f\\ \hline 
\ea$$

We will be interested in the following gauge invariant composite operators 
that can be built from the
elementary fields:
\bea \label{mesbar2adj}
     mesons  &&M_{(j,l)}\equiv Q Q_{(j,l)},
               \quad\w Q_{(j,l)}=X^jY^l Q,
               \quad j=0,\ldots,k-1,\ l=0,1,2,\nonumber\\
     baryons &&B^{(n_{(0,0)},\ldots,n_{(k-1,2)})}\equiv 
               Q^{n_{(0,0)}}Q_{(1,0)}^{n_{(1,0)}}\cdots
               Q_{(k-1,2)}^{n_{(k-1,2)}},\nonumber\\ 
             &&\bar B^{(n_{(0,0)},\ldots,n_{(k-1,2)})}\equiv 
               \bar Q^{n_{(0,0)}}\bar Q_{(1,0)}^{n_{(1,0)}}\cdots
               \bar Q_{(k-1,2)}^{n_{(k-1,2)}},\\ 
             &&\qquad\w\sum_{l=0}^2\sum_{j=0}^{k-1}n_{(j,l)}=N_c,\nonumber
\eea
where the gauge indices are contracted with an epsilon tensor for the baryons. 

For odd values of $k$ this theory is dual to an $SU(\tilde N_c)$ gauge theory,
with $\tilde N_c=3kN_f-N_c$ and matter content \cite{Bro}
$$\ba{|c|c|cccccc|}
\hline  \rule[-1.3ex]{0mm}{4ex}
& SU(\tilde N_c) & SU(N_f)_L & SU(N_f)_R & U(1)_B 
& U(1)_R & {\bfm Z}_{2(k+1)N_f} & {\bfm Z}_{2N_f} \\ \hline
\rule{0mm}{3ex}
q  &\Yfun &\Yfunb &\bfm 1 &{1\over\tilde Nc} 
          &1-{\tilde N_c\over(k+1)N_f} &-\tilde N_c &-\tilde N_c\\
\bar q  &\Yfunb &\bfm 1 &\Yfunb &-{1\over\tilde Nc} 
                &1-{\tilde N_c\over(k+1)N_f} &-\tilde N_c &-\tilde N_c\\ 
\tilde X  &adj &\bfm 1 &\bfm 1 &0 &{2\over k+1} &2N_f &0\\ 
\rule[-1.5ex]{0mm}{1.5ex}
\tilde Y  &adj &\bfm 1 &\bfm 1 &0 &{k\over k+1} &-N_f &N_f\\ \hline
\ea$$
and singlets $M_{(j,l)}$ that carry the same quantum numbers
as the mesons of the electric theory.

The magnetic theory contains a tree-level superpotential
\be \label{Wmag2adj}
    \Wmag=\Tr\tilde X^{k+1}\ +\ \Tr\tilde X\tilde Y^2\ 
          +\ \sum_{l=0}^2\sum_{j=0}^{k-1}M_{(k-1-j,2-l)}q\bar q_{(j,l)},
\ee
where magnetic dressed quarks have been introduced by 
$q_{(j,l)}=\tilde X^j\tilde Y^l q$ and we have omitted the dependence of
$\Wmag$ on $h_1$, $h_2$ and $\mu$.

The electric baryons of (\ref{mesbar2adj}) can be consistently mapped to 
similar baryons of the magnetic theory \cite{Bro}:
\be \label{barmap2adj}
   B^{(n_{(0,0)},\ldots,n_{(k-1,2)})}\ \lra\ 
   \widetilde B^{(m_{(0,0)},\ldots,m_{(k-1,2)})},
   \quad\w\ m_{(j,l)}=N_f-n_{(k-1-j,2-l)}, 
\ee
where the magnetic baryons $\widetilde B^{(\ldots)}$
are defined in the same way as the electric baryons of (\ref{mesbar2adj}) 
replacing all fields by their dual partners and $N_c$ by $\tilde N_c$. 

For $N_c=3kN_f-1$ the magnetic theory is completely higgsed and the 
electric theory confines with low-energy spectrum given by 
the composite fields 
\bea \label{confspec2adj}
     &&M_{(j,l)},\quad j=0,\ldots,k-1,\ l=0,1,2,\nonumber\\
     &&B\equiv B^{(N_f,\ldots,N_f,N_f-1)},\quad 
       \bar B\equiv \bar B^{(N_f,\ldots,N_f,N_f-1)}
\eea 
of eqs.\ (\ref{mesbar2adj}). From (\ref{barmap2adj}) we find that 
the baryons are mapped to the $2N_f$ magnetic quark singlets $q$, $\bar q$
that stay massless after breaking the magnetic gauge group. The 't Hooft 
anomaly matching conditions are satisfied because the only fields that 
contribute to the global anomalies in the magnetic theory for $\tilde N_c=1$ 
are the $2N_f$ quark singlets and the meson singlets which carry the same 
charges as the baryons and mesons of the electric theory.

The effective low-energy superpotential of the magnetic theory, deduced
from (\ref{Wmag2adj}) for the theory with $\tilde N_c=3k+1$ and $(N_f+1)$
quark flavors by adding a tree-level term $mM_{(0,0)}$ and integrating out 
the massive modes, contains a term $M_{(k-1,2)}q\bar q$. We thus expect that 
the confining superpotential of the electric theory has a term proportional to
\be \bar B M_{k-1} B\ \sim\ (Q\bar Q)^{3kN_f}X^{(k-1)N_c}Y^{2N_c}.
\ee
Comparing this to (\ref{Weff2}) we find $\alpha=3k$. From (\ref{albega2}) 
we then get $\gamma_1=-2N_c$ and $\gamma_2=-(3k-1)N_c$. 
The confining superpotential consequently is of the form
\bea \label{Wconf2adj}
    W &= &{\bar B M_{(k-1,2)} B\ +\sum_{\{j_m,l_m\}}(\det M_{(k-1,2)})^{3k-p}
      \prod_{l=1}^p(M_{(j_m,l_m)}\cof M_{(k-1,2)})
      \over h_1^{2N_c}h_2^{(3k-1)N_c}\,\Lambda^{3k((3k-1)N_f-1)}},\\
    \w && \sum_{m=1}^pj_m=(k-1)(p-1),\quad\sum_{m=1}^pl_m=2(p-1),\nonumber
\eea
where the cofactor of a matrix $A$ is defined by $(\cof A)^{ij}=\partial\det A
/\partial A_{ij}$. To determine the classical constraints we view the $3k$
dressed quarks $\cQ=(Q,Q_{(1,0)},\ldots,Q_{(k-1,2)})$ as independent degrees
of freedom \cite{newconf}. We restrict ourselves to the special case where
the matrices $X$ and $Y$ commute (in general the constraints are more 
complicated). The mesons $\cM=\cQ\bar\cQ$ and the baryons $\cB=\cQ^{N_c}$,
$\bar\cB=\bar\cQ^{N_c}$ should then satisfy the constraints that are known
for SQCD with one more flavor than colors:
\be \label{constr2adj}
    \cM\cB=\bar\cB\cM=0\qquad{\rm and}\qquad(\cof\cM)^{ij}=\cB^i\bar\cB^j.
\ee
The first condition reduces to $M_{(k-1,2)}B=\bar BM_{(k-1,2)}=0$ because
the baryons $\cB$, $\bar\cB$ are given by \cite{newconf} $\cB=(0,\ldots,B)$,
$\bar\cB=(0,\ldots,\bar B)$. These constraints are correctly reproduced by 
the equations of motion of the superpotential (\ref{Wconf2adj}).

In the limit $k=1$ the coupling parameter $h_1$ represents a mass for the
tensor $X$. Integrating out $X$ results in a theory with tree-level
superpotential $\Wtree=h\,\Tr Y^4$, where $h=(h_2)^2/h_1$, as explained
in \cite{Bro}. This is the model of Kutasov and Schwimmer \cite{KuSch}
(reviewed in section 2 of the present paper) for the special case $k=3$.
For $N_c=3N_f-1$ it confines \cite{KSS,newconf}. The confining superpotential
(\ref{Wconf2adj}) has the correct form 
$W=\bar B M_{(0,2)} B+\ldots/h^{N_c}\Lambda_L^{3(5N_f-2)}$ in this limit, 
where we used the scale matching relation 
$h_1^{N_c}\Lambda^{N_c-N_f}=\Lambda_L^{2N_c-N_f}$.

\subsection{$SU(N_c)$ with an adjoint tensor, an \asym tensor and its 
            conjugate}
Consider an $SU(N_c)$ gauge theory with $N_f$ fundamental flavors $Q$, 
$\bar Q$, an adjoint tensor $X$, an \asym tensor $Y$ and its conjugate 
$\bar Y$ and tree-level superpotential 
$\Wtree=h_1\,\Tr X^{k+1}\ +\ h_2\,\Tr XY\bar Y$. 
This model was first studied in \cite{patterns}. The transformation
properties of the matter fields under the gauge symmetry and the \nonan
global symmetries are shown in the following table:
$$\ba{|c|c|ccccccc|}
\hline   \rule[-1.3ex]{0mm}{4ex}
& SU(N_c) & SU(N_f)_L & SU(N_f)_R &U(1)_Y & U(1)_B 
& U(1)_R & {\bfm Z}_{2(k+1)N_f} & {\bfm Z}_{2N_f} \\ \hline
\rule{0mm}{3ex}
Q  &\Yfun &\Yfun &\bfm 1 &0 &{1\over N_c} &1-{N_c+2\over(k+1)N_f} 
          &-(N_c+2) &-(N_c-2)\\
\bar Q  &\Yfunb &\bfm 1 &\Yfun &0 &-{1\over N_c} &1-{N_c+2\over(k+1)N_f} 
                &-(N_c+2) &-(N_c-2)\\ 
X  &adj &\bfm 1 &\bfm 1 &0 &0 &{2\over k+1} &2N_f &0\\
Y  &\Yasym &\bfm 1 &\bfm 1 &1 &{2\over N_c} &{k\over k+1} &-N_f &N_f\\
\rule[-1.5ex]{0mm}{1.5ex}
\bar Y  &\Yasymb &\bfm 1 &\bfm 1 &-1 &-{2\over N_c} &{k\over k+1} &-N_f &N_f\\ 
\hline 
\ea$$

We will be interested in the following gauge invariant composite operators 
that can be built from the
elementary fields:
\bea \label{mesbaradas}
mesons && M_j=Q\bar Q_{(j)},\quad N_j=Q\bar YY\bar Q_{(j)},\quad
   P_j= Q\bar YQ_{(j)},\quad \bar P_j=\bar QY\bar Q_{(j)},\nonumber\\
   &&\qquad\w\ Q_{(j)}=X^j Q,\ \bar Q_{(j)}=X^j\bar Q,\quad 
     j=0,\ldots,k-1,\nonumber \\
baryons && B_n^{(n_0,\ldots,n_{k-1})}=Y^n Q^{n_0}Q_{(1)}^{n_1}\cdots
           Q_{(k-1)}^{n_{k-1}},\\
        && \bar B_n^{(n_0,\ldots,n_{k-1})}=\bar Y^n \bar Q^{n_0}
           \bar Q_{(1)}^{n_1}\cdots\bar Q_{(k-1)}^{n_{k-1}},\nonumber\\
        && \qquad\w \sum_{j=0}^{k-1}n_j=N_c-2n,\nonumber
\eea
where the gauge indices of the baryons are contracted with an epsilon tensor.

For odd values of $k$ this theory is dual to an $SU(\tilde N_c)$ gauge theory,
with $\tilde N_c=3kN_f-4-N_c$ and matter content \cite{patterns}
$$\ba{|c|c|ccccccc|}
\hline   \rule[-1.3ex]{0mm}{4ex}
& SU(\tilde N_c) & SU(N_f)_L & SU(N_f)_R &U(1)_Y & U(1)_B 
& U(1)_R & {\bfm Z}_{2(k+1)N_f} & {\bfm Z}_{2N_f} \\ \hline
\rule{0mm}{3ex}
q  &\Yfun &\Yfunb &\bfm 1 &{kN_f-2\over\tilde N_c} &{1\over\tilde N_c} 
          &1-{\tilde N_c+2\over(k+1)N_f} &-(\tilde N_c+2) &-(\tilde N_c+6)\\
\bar q  &\Yfunb &\bfm 1 &\Yfunb &-{kN_f-2\over\tilde N_c} 
          &-{1\over\tilde N_c} &1-{\tilde N_c+2\over(k+1)N_f} 
          &-(\tilde N_c+2) &-(\tilde N_c+6)\\ 
\tilde X &adj &\bfm 1 &\bfm 1 &0 &0 &{2\over k+1} &2N_f &0\\
\tilde Y &\Yasym &\bfm 1 &\bfm 1 &{N_c-kN_f\over\tilde N_c} &{2\over\tilde N_c}
                 &{k\over k+1} &-N_f &N_f\\
\rule[-1.5ex]{0mm}{1.5ex}
\tilde{\bar Y} &\Yasymb &\bfm 1 &\bfm 1 &-{N_c-kN_f\over\tilde N_c} 
               &-{2\over\tilde N_c} &{k\over k+1} &-N_f &N_f\\ \hline
\ea$$
and singlets $M_j$, $N_j$, $P_j$, $\bar P_j$ that carry the same quantum 
numbers as the mesons of the electric theory.

The magnetic theory contains a tree-level superpotential
\be \label{Wmagadas}
   \Wmag =\Tr \tilde X^{k+1}+\Tr\tilde X\tilde Y\tilde{\bar Y}
          +\sum_{j=0}^{k-1} \Big[M_{k-1-j} q\tilde{\bar Y}\tilde Y\bar q_{(j)}
          +N_{k-1-j} q\bar q_{(j)}+P_{k-1-j} q\tilde{\bar Y} q_{(j)}
                    +\bar P_{k-1-j}\bar q\tilde Y\bar q_{(j)}\Big]\ ,
\ee
where magnetic dressed quarks have been introduced by $q_{(j)}=\tilde X^j q$.

The authors of \cite{patterns} found a mapping of the baryons $B_n^{(\cdots)}$
of (\ref{mesbaradas}) to the magnetic baryons 
$\widetilde B_m^{(m_0,\ldots,m_{k-1})}=\tilde Y^mq^{m_0}q_{(1)}^{m_1}
\cdots q_{(k-1)}^{m_{k-1}}$ consistent with all global symmetries:
\be \label{barmapadas}
    B_n^{(n_0,\ldots,n_{k-1})} \lra \widetilde B_m^{(m_0,\ldots,m_{k-1})},
    \quad \w\ m=kN_f-2-n,\quad m_j=N_f-n_{k-1-j}.
\ee

For $N_c=3kN_f-5$ the magnetic theory is completely higgsed and the 
electric theory confines with low-energy spectrum given by 
the composite fields 
\bea \label{confspecadas}
&&M_j,\ N_j,\ P_j,\  \bar P_j,\quad j=0,\ldots,k-1,\nonumber\\
&&B\equiv B_{kN_f-2}^{(N_f,\ldots,N_f,N_f-1)},\ 
  \bar B\equiv \bar B_{kN_f-2}^{(N_f,\ldots,N_f,N_f-1)}
\eea 
of eqs.\ (\ref{mesbaradas}). From (\ref{barmapadas}) we find that 
the baryons are mapped to the $2N_f$ magnetic quark singlets $q$, $\bar q$
that stay massless after breaking the magnetic gauge group. The 't Hooft 
anomaly matching conditions are satisfied because the only fields that 
contribute to the global anomalies in the magnetic theory for $\tilde N_c=1$ 
are the $2N_f$ quark singlets and the meson singlets which carry the same 
charges as the baryons and mesons of the electric theory.

The effective low-energy superpotential of the magnetic theory, deduced
from (\ref{Wmagadas}) for the theory with $\tilde N_c=3k+1$ and $(N_f+1)$
quark flavors by adding a tree-level term $mM_{0}$ and integrating out 
the massive modes, contains a term $N_{k-1}q\bar q$. We thus expect that 
the confining superpotential of the electric theory has a term proportional to
\be \bar B N_{k-1} B\ \sim\ 
    (Q\bar Q)^{kN_f}X^{(k-1)(kN_f-1)}(Y\bar Y)^{kN_f-1}.
\ee
Comparing this to (\ref{Weff2}) we find $\alpha=k$. From (\ref{albega2}) 
we then get $\gamma_1=-2(kN_f-1)$ and $\gamma_2=-k(3k-1)N_f+7k-1$. 
The confining superpotential consequently is of the form
\be \label{Wconfadas}
    W={\bar B N_{k-1} B\over h_1^{2(kN_f-1)}h_2^{k(3k-1)N_f-7k+1}\,
      \Lambda^{k((3k-1)N_f-3)}}\ +\ \ldots,
\ee
where the dots stand for products of $M_j$, $N_j$ and $(P_j\bar P_j)$
that are invariant under the symmetries. These terms may contain different
powers of the fields $X$ and $Y$.

In the limit $k=1$ the coupling parameter $h_1$ represents a mass for the
tensor $X$. Integrating out $X$ results in a theory with tree-level
superpotential $\Wtree=h\,\Tr(Y\bar Y)^2$, where $h=(h_2)^2/h_1$. This is 
the model discussed in section 3.1 for the special case $k=1$.
For $N_c=3N_f-5$ it confines \cite{newconf}. The confining superpotential
(\ref{Wconfadas}) has the correct form 
$W=(\bar B N_0 B+\ldots)/h^{N_f-3}\Lambda_L^{5N_f-8}$ in this limit, 
where we used the scale matching relation 
$h_1^{N_c}\Lambda^{N_c+2-N_f}=\Lambda_L^{2N_c+2-N_f}$.

\subsection{$SU(N_c)$ with an adjoint tensor, a symmetric tensor and its 
            conjugate}
Consider an $SU(N_c)$ gauge theory with $N_f$ fundamental flavors $Q$, 
$\bar Q$, an adjoint tensor $X$, a symmetric tensor $Y$ and its conjugate 
$\bar Y$ and tree-level superpotential 
$\Wtree=h_1\,\Tr X^{k+1}\ +\ h_2\,\Tr XY\bar Y$. 
This model was first studied in \cite{patterns}. The transformation
properties of the matter fields under the gauge symmetry and the \nonan
global symmetries are shown in the following table:
$$\ba{|c|c|ccccccc|}
\hline   \rule[-1.3ex]{0mm}{4ex}
& SU(N_c) & SU(N_f)_L & SU(N_f)_R &U(1)_Y & U(1)_B 
& U(1)_R & {\bfm Z}_{2(k+1)N_f} & {\bfm Z}_{2N_f} \\ \hline
\rule{0mm}{3ex}
Q  &\Yfun &\Yfun &\bfm 1 &0 &{1\over N_c} &1-{N_c-2\over(k+1)N_f} 
          &-(N_c-2) &-(N_c+2)\\
\bar Q  &\Yfunb &\bfm 1 &\Yfun &0 &-{1\over N_c} &1-{N_c-2\over(k+1)N_f} 
                &-(N_c-2) &-(N_c+2)\\ 
X  &adj &\bfm 1 &\bfm 1 &0 &0 &{2\over k+1} &2N_f &0\\
Y  &\Ysym &\bfm 1 &\bfm 1 &1 &{2\over N_c} &{k\over k+1} &-N_f &N_f\\
\rule[-1.5ex]{0mm}{1.5ex}
\bar Y  &\Ysymb &\bfm 1 &\bfm 1 &-1 &-{2\over N_c} &{k\over k+1} &-N_f &N_f\\ 
\hline 
\ea$$

We will be interested in the following gauge invariant composite operators 
that can be built from the elementary fields:
\bea \label{mesbaradsy}
mesons && M_j=Q\bar Q_{(j)},\quad N_j=Q\bar YY\bar Q_{(j)},\quad
   P_j= Q\bar YQ_{(j)},\quad \bar P_j=\bar QY\bar Q_{(j)},\nonumber\\
   &&\qquad\w\ Q_{(j)}=X^j Q,\ \bar Q_{(j)}=X^j\bar Q,\quad 
     j=0,\ldots,k-1,\nonumber \\
baryons && \cB^{(n_0,\ldots,n_{k-1},\bar n_0,\ldots,\bar n_{k-1},
              \bar{\bar n}_0,\ldots,\bar{\bar n}_{k-1})} =(YX^{k-1}W_\alpha)^2
           Q^{n_0}Q_{(1)}^{n_1}\cdots Q_{(k-1)}^{n_{k-1}}\nonumber\\
        && \qquad\cdot(Y\bar Q)^{\bar n_0}(Y\bar Q_{(1)})^{\bar n_1}\cdots
           (Y\bar Q_{(k-1)})^{\bar n_{k-1}} 
           (Y\bar YQ)^{\bar{\bar n}_0}(Y\bar YQ_{(1)})^{\bar{\bar n}_1}
           \cdots(Y\bar YQ_{(k-1)})^{\bar{\bar n}_{k-1}},\nonumber\\
        && \bar\cB^{(n_0,\ldots,n_{k-1},\bar n_0,\ldots,\bar n_{k-1},
               \bar{\bar n}_0,\ldots,\bar{\bar n}_{k-1})} 
           =(\bar YX^{k-1}W_\alpha)^2 \bar Q^{n_0}\bar Q_{(1)}^{n_1}\cdots 
           \bar Q_{(k-1)}^{n_{k-1}}\nonumber\\
        && \qquad\cdot(\bar YQ)^{\bar n_0}(\bar YQ_{(1)})^{\bar n_1}\cdots
           (\bar YQ_{(k-1)})^{\bar n_{k-1}} (\bar YY\bar Q)^{\bar{\bar n}_0}
           (\bar YY\bar Q_{(1)})^{\bar{\bar n}_1}\cdots
           (\bar YY\bar Q_{(k-1)})^{\bar{\bar n}_{k-1}},\nonumber\\
        && \qquad\w \sum_{j=0}^{k-1}\left(n_j+\bar n_j+\bar{\bar n}_j\right)
            =N_c-4,\\
        && B_n^{(n_0,\ldots,n_{k-1})}=Y^n Q^{n_0}Q^{n_0}
           Q_{(1)}^{n_1}Q_{(1)}^{n_1}\cdots
           Q_{(k-1)}^{n_{k-1}}Q_{(k-1)}^{n_{k-1}},\nonumber\\
        && \bar B_n^{(n_0,\ldots,n_{k-1})}=\bar Y^n \bar Q^{n_0}\bar Q^{n_0}
           \bar Q_{(1)}^{n_1}\bar Q_{(1)}^{n_1}\cdots
           \bar Q_{(k-1)}^{n_{k-1}}\bar Q_{(k-1)}^{n_{k-1}},\nonumber\\
        && \qquad\w \sum_{j=0}^{k-1}n_j=N_c-n,\nonumber
\eea
where the gauge indices of the baryons are contracted with one epsilon tensor 
for the $\cB^{(\cdots)}$, $\bar\cB^{(\cdots)}$ and with two epsilon tensors 
for the $B_n^{(\cdots)}$, $\bar B_n^{(\cdots)}$.

For odd values of $k$ this theory is dual to an $SU(\tilde N_c)$ gauge theory,
with $\tilde N_c=3kN_f+4-N_c$ and matter content \cite{patterns}
$$\ba{|c|c|ccccccc|}
\hline   \rule[-1.3ex]{0mm}{4ex}
& SU(\tilde N_c) & SU(N_f)_L & SU(N_f)_R &U(1)_Y & U(1)_B 
& U(1)_R & {\bfm Z}_{2(k+1)N_f} & {\bfm Z}_{2N_f} \\ \hline
\rule{0mm}{3ex}
q  &\Yfun &\Yfunb &\bfm 1 &{kN_f+2\over\tilde N_c} &{1\over\tilde N_c} 
          &1-{\tilde N_c-2\over(k+1)N_f} &-(\tilde N_c-2) &-(\tilde N_c-6)\\
\bar q  &\Yfunb &\bfm 1 &\Yfunb &-{kN_f+2\over\tilde N_c} 
          &-{1\over\tilde N_c} &1-{\tilde N_c-2\over(k+1)N_f} 
          &-(\tilde N_c-2) &-(\tilde N_c-6)\\ 
\tilde X &adj &\bfm 1 &\bfm 1 &0 &0 &{2\over k+1} &2N_f &0\\
\tilde Y &\Ysym &\bfm 1 &\bfm 1 &{N_c-kN_f\over\tilde N_c} &{2\over\tilde N_c}
                 &{k\over k+1} &-N_f &N_f\\
\rule[-1.5ex]{0mm}{1.5ex}
\tilde{\bar Y} &\Ysymb &\bfm 1 &\bfm 1 &-{N_c-kN_f\over\tilde N_c} 
               &-{2\over\tilde N_c} &{k\over k+1} &-N_f &N_f\\ \hline
\ea$$
and singlets $M_j$, $N_j$, $P_j$, $\bar P_j$ that carry the same quantum 
numbers as the mesons of the electric theory.

The magnetic theory contains a tree-level superpotential
\be \label{Wmagadsy}
   \Wmag =\Tr \tilde X^{k+1}+\Tr\tilde X\tilde Y\tilde{\bar Y}
          +\sum_{j=0}^{k-1} \Big[M_{k-1-j} q\tilde{\bar Y}\tilde Y\bar q_{(j)}
          +N_{k-1-j} q\bar q_{(j)}+P_{k-1-j} q\tilde{\bar Y} q_{(j)}
                    +\bar P_{k-1-j}\bar q\tilde Y\bar q_{(j)}\Big]\ ,
\ee
where $q_{(j)}=\tilde X^j q$.

The electric baryons $\cB^{(\cdots)}$ and $B_n^{(\cdots)}$ of 
(\ref{mesbaradsy}) can be consistently mapped to the magnetic baryons 
$\widetilde\cB^{(m_i,\bar m_j,\bar{\bar m}_l)}= q^{m_0}\cdots 
q_{(k-1)}^{m_{k-1}}(\tilde Y\bar q)^{\bar m_0}\cdots
(\tilde Y\bar q_{(k-1)})^{\bar m_{k-1}} (\tilde Y\tilde{\bar Y}q)^{
\bar{\bar m}_0}\cdots(\tilde Y\tilde{\bar Y}q_{(k-1)})^{\bar{\bar m}_{k-1}}$
and $\widetilde B_m^{(m_i)}=Y^mq^{m_0}q^{m_0}\cdots q_{(k-1)}^{m_{k-1}}
q_{(k-1)}^{m_{k-1}}$ according to the following prescription:
\bea \label{barmapadsy}
   \cB^{(n_0,\ldots,n_{k-1},\bar n_0,\ldots,\bar n_{k-1},
         \bar{\bar n}_0,\ldots,\bar{\bar n}_{k-1})} 
   &\lra 
   &\widetilde\cB^{(m_0,\ldots,m_{k-1},\bar m_0,\ldots,\bar m_{k-1},
       \bar{\bar m}_0,\ldots,\bar{\bar m}_{k-1})}, \quad\w\nonumber\\
   m_j=N_f-\bar{\bar n}_{k-1-j}, &&\bar m_j=N_f-\bar n_{k-1-j},\quad
     \bar{\bar m}_j=N_f-n_{k-1-j},\nonumber\\ 
  B_n^{(n_0,\ldots,n_{k-1})} &\lra &\widetilde B_m^{(m_0,\ldots,m_{k-1})},
      \quad \w\\
   m=2kN_f+4-n, &&m_j=N_f-n_{k-1-j}.\nonumber
\eea
The last of these mappings has been found in \cite{patterns}.

For $N_c=3kN_f+3$ the magnetic theory is completely higgsed and the 
electric theory confines with low-energy spectrum given by 
the composite fields 
\bea \label{confspecadsy}
&&M_j,\ N_j,\ P_j,\  \bar P_j,\quad j=0,\ldots,k-1,\nonumber\\
&&B\equiv \cB^{(N_f,\ldots,N_f,N_f-1)},\ 
  \bar B\equiv \bar\cB^{(N_f,\ldots,N_f,N_f-1)},\\ 
&&b\equiv B_{2kN_f+3}^{(N_f,\ldots,N_f)},\ 
  \bar b\equiv \bar B_{2kN_f+3}^{(N_f,\ldots,N_f)}\nonumber
\eea 
of eqs.\ (\ref{mesbaradsy}). From (\ref{barmapadsy}) we find the mappings 
$B,\bar B\lra q,\bar q$ and $b,\bar b\lra Y,\bar Y$. One color component of 
each of the fields $q$, $\bar q$, $Y$, $\bar Y$ together with the meson 
singlets are exactly the degrees of 
freedom that stay massless after breaking the magnetic gauge group. The 
't Hooft anomaly matching conditions are satisfied because the only fields 
that contribute to the global anomalies in the magnetic theory for 
$\tilde N_c=1$ are the $2N_f$ quark singlets, one component of $Y$ and 
$\bar Y$ each and the meson singlets.

The effective low-energy superpotential of the magnetic theory, deduced
from (\ref{Wmagadsy}) for the theory with $\tilde N_c=3k+1$ and $(N_f+1)$
quark flavors by adding a tree-level term $mM_{0}$ and integrating out 
the massive modes, contains the terms 
$$N_{k-1}q\bar q\ +\ P_{k-1}q\tilde{\bar Y}q\ +\ 
                     \bar P_{k-1}\bar q\tilde Y\bar q.$$ 
We thus expect that the confining superpotential of the electric theory has 
terms proportional to
\bea  \label{conftermsadsy}
    \bar B N_{k-1} B &\sim 
    &(W_\alpha)^4(Q\bar Q)^{3kN_f}X^{(k-1)N_c}(Y\bar Y)^{3kN_f+1},\nonumber\\
    BBP_{k-1}\bar b,\ \bar B\bar B\bar P_{k-1}b &\sim 
    &(W_\alpha)^4(Q\bar Q)^{4kN_f}X^{(k-1)(N_c+kN_f)}(Y\bar Y)^{4kN_f+2}.
\eea
Comparing this to (\ref{Weff2}) we find $\alpha^{(1)}=3k$ for the term in the
first line of (\ref{conftermsadsy}) and from (\ref{albega2}, \ref{albega3}) 
$\gamma^{(1)}_1=-6kN_f-2$ and $\gamma^{(1)}_2=-3k(3k-1)N_f-15k+1$. 
For the terms in the second line we get $\alpha^{(2)}=4k$ and
$\gamma^{(2)}_1=-8kN_f-3$ and $\gamma^{(2)}_2=-4k(3k-1)N_f-20k+2$.
The confining superpotential consequently is of the form
\bea \label{Wconfadsy}
    W &= &{\bar B N_{k-1} B\over h_1^{6kN_f+2}h_2^{3k(3k-1)N_f+15k-1}\,
      \Lambda^{3k((3k-1)N_f+1)}}\nonumber\\ 
      &&+\ {BBP_{k-1}\bar b + \bar B\bar B\bar P_{k-1}b\over 
       h_1^{8kN_f+3}h_2^{4k(3k-1)N_f+20k-2}\,
      \Lambda^{4k((3k-1)N_f+1)}}\ +\ \ldots.
\eea
Again there may be further terms generated by instantons of the magnetic
gauge group.

In the limit $k=1$ the coupling parameter $h_1$ represents a mass for the
tensor $X$. Integrating out $X$ results in a theory with tree-level
superpotential $\Wtree=h\,\Tr(Y\bar Y)^2$, where $h=(h_2)^2/h_1$. This is 
the model discussed in section 3.2 for the special case $k=1$.
For $N_c=3N_f+3$ it confines \cite{newconf}. The  confining superpotential
(\ref{Wconfadsy}) contains an additional term proportional to $(b\bar b)^2$
for $k=1$ and has the correct form 
$$W={\bar B N_0 B\over h^{3N_f+7}\Lambda_L^{3(5N_f+4)}}\ +\ 
    {(b\bar b)^2+h^{-2}(BBP_0\bar b+\bar B\bar B\bar P_0b)\over
    h^{4N_f+7}\Lambda^{4(5N_f+4)}}$$ 
in this limit, where we used the scale matching relation 
$h_1^{N_c}\Lambda^{N_c-2-N_f}=\Lambda_L^{2N_c-2-N_f}$.

\subsection{$SU(N_c)$ with an adjoint tensor, an \asym tensor and a 
            conjugate symmetric tensor}
Consider an $SU(N_c)$ gauge theory with $N_f+8$ quarks $Q$, $N_f$ 
antiquarks $\bar Q$,  an adjoint tensor $X$, an \asym tensor $Y$ and 
a conjugate symmetric tensor $\bar Y$ and tree-level superpotential 
$\Wtree=h_1\,\Tr X^{k+1}\ +\ h_2\,\Tr XY\bar Y$. 
This model was first studied in \cite{patterns}. The transformation
properties of the matter fields under the gauge symmetry and the \nonan
global symmetries are shown in the following table:
$$\ba{|c|c|ccccccc|}
\hline   \rule[-1.3ex]{0mm}{4ex}
& SU(N_c) & \msmall{SU(N_f+8)} &\!\msmall{SU(N_f)} &U(1)_Y & U(1)_B 
& U(1)_R & {\bfm Z}_{2(k+1)(N_f+4)} & {\bfm Z}_{2(N_f+4)} \\ \hline
\rule{0mm}{3ex}
Q  &\Yfun &\Yfun &\bfm 1 &{6\over N_f+8}\msmall{-1} &{1\over N_c} 
          &\msmall{1-}{N_c+6k\over(k+1)(N_f+8)} &-N_c &-N_c\\
\bar Q  &\Yfunb &\bfm 1 &\Yfun &{6\over N_f}\msmall{+1} &-{1\over N_c} 
                &\msmall{1-}{N_c-6k\over(k+1)N_f} &-N_c &-N_c\\ 
X  &adj &\bfm 1 &\bfm 1 &0 &0 &{2\over k+1} &2(N_f+4) &0\\
Y  &\Yasym &\bfm 1 &\bfm 1 &1 &{2\over N_c} &{k\over k+1} &-(N_f+4) &N_f+4\\
\rule[-1.5ex]{0mm}{1.5ex}
\bar Y  &\Ysymb &\bfm 1 &\bfm 1 &-1 &-{2\over N_c} &{k\over k+1} 
                &-(N_f+4) &N_f+4\\ 
\hline 
\ea$$

We will be interested in the following gauge invariant composite operators 
that can be built from the
elementary fields:
\bea \label{mesbaradsas}
mesons && M_j=Q\bar Q_{(j)},\quad N_j=Q\bar YY\bar Q_{(j)},\quad
   P_j= Q\bar YQ_{(j)},\quad \bar P_j=\bar QY\bar Q_{(j)},\nonumber\\
   &&\qquad\w\ Q_{(j)}=X^j Q,\ \bar Q_{(j)}=X^j\bar Q,\quad 
     j=0,\ldots,k-1,\nonumber \\
baryons && \bar\cB_p^{(n_0,\ldots,n_{k-1},\bar n_0,\ldots,\bar n_{k-1},
               \bar{\bar n}_0,\ldots,\bar{\bar n}_{k-1})} 
           =(\bar YW_\alpha)^2(\bar YXW_\alpha)^2\cdots
            (\bar YX^{p-1}W_\alpha)^2\nonumber\\ 
        && \hskip3cm \cdot\bar Q^{n_0}\bar Q_{(1)}^{n_1}\cdots 
           \bar Q_{(k-1)}^{n_{k-1}} 
           (\bar YQ)^{\bar n_0}(\bar YQ_{(1)})^{\bar n_1}\cdots
           (\bar YQ_{(k-1)})^{\bar n_{k-1}} \nonumber\\
        && \hskip3cm \cdot(\bar YY\bar Q)^{\bar{\bar n}_0}
           (\bar YY\bar Q_{(1)})^{\bar{\bar n}_1}\cdots
           (\bar YY\bar Q_{(k-1)})^{\bar{\bar n}_{k-1}},\nonumber\\
        && \qquad\w \sum_{j=0}^{k-1}\left(n_j+\bar n_j+\bar{\bar n}_j\right)
            =N_c-4p,\nonumber\\
        && B_n^{(n_0,\ldots,n_{k-1})}=Y^n Q^{n_0}Q_{(1)}^{n_1}\cdots
           Q_{(k-1)}^{n_{k-1}},\\
        && \bar B_{\bar n}^{(\bar n_0,\ldots,\bar n_{k-1})}=\bar Y^{\bar n} 
           \bar Q^{\bar n_0}\bar Q^{\bar n_0} \bar Q_{(1)}^{\bar n_1}
           \bar Q_{(1)}^{\bar n_1}\cdots\bar Q_{(k-1)}^{\bar n_{k-1}}
           \bar Q_{(k-1)}^{\bar n_{k-1}},\nonumber\\
        && \qquad\w \sum_{j=0}^{k-1}n_j=N_c-2n,\quad
            \sum_{j=0}^{k-1}\bar n_j=N_c-\bar n,\nonumber
\eea
where the gauge indices of the baryons are contracted with one epsilon tensor 
for the $\bar\cB_p^{(\cdots)}$, $B_n^{(\cdots)}$ and with two epsilon tensors 
for the $\bar B_{\bar n}^{(\cdots)}$.

This theory is dual to an $SU(\tilde N_c)$ gauge theory, with 
$\tilde N_c=3k(N_f+4)-N_c$ and matter content \cite{patterns}\footnote{%
We corrected some misprints in the $U(1)_Y$ charges associated to the matter 
fields in \cite{patterns}.}
$$\ba{|c|c|ccccccc|}
\hline   \rule[-1.3ex]{0mm}{4ex}
& SU(\tilde N_c) &\msmall{SU(N_f+8)} &\!\msmall{SU(N_f)} &U(1)_Y & U(1)_B 
& U(1)_R &\!{\bfm Z}_{2(k+1)(N_f+4)} & {\bfm Z}_{2(N_f+4)} \\ \hline
\rule{0mm}{3ex}
q  &\Yfun &\Yfunb &\bfm 1 &\msmall{1-}{6\over N_f+8} &{1\over\tilde N_c} 
          &\msmall{1-}{\tilde N_c+6k\over(k+1)(N_f+8)} 
          &\msmall{-\tilde N_c-p_1} &\msmall{-\tilde N_c-p_2}\\
\bar q  &\Yfunb &\bfm 1 &\Yfunb &\msmall{-1-}{6\over N_f}  
          &-{1\over\tilde N_c} &\msmall{1-}{\tilde N_c-6k\over(k+1)N_f} 
          &\msmall{-\tilde N_c+p_1} &\msmall{-\tilde N_c+p_2}\\
\tilde X &adj &\bfm 1 &\bfm 1 &\msmall{0} &\msmall{0} &{2\over k+1} 
              &\msmall{2(N_f+4)} &\msmall{0}\\
\tilde Y &\Yasym &\bfm 1 &\bfm 1 &\msmall{-1} &{2\over\tilde N_c} 
                 &{k\over k+1} &\msmall{-(N_f+4)-2p_1} &\msmall{N_f+4-2p_2}\\
\rule[-1.5ex]{0mm}{1.5ex}
\tilde{\bar Y} &\Ysymb &\bfm 1 &\bfm 1 &\msmall{1} &-{2\over\tilde N_c} 
                       &{k\over k+1} &\msmall{-(N_f+4)+2p_1} 
                       &\msmall{N_f+4+2p_2}\\ \hline
\ea$$
and singlets $M_j$, $N_j$, $P_j$, $\bar P_j$ that carry the same quantum 
numbers as the mesons of the electric theory. The numbers $p_1$, $p_2$ are
defined by 
$$p_1=2+{2(2k+1)N_c\over\tilde N_c},\qquad p_2={4kN_c\over\tilde N_c}.$$
They are determined by requiring that the baryon mappings below be invariant
under the discrete symmetries. We checked that the discrete anomaly
matching conditions are satisfied with these charges for the matter fields.

The magnetic theory contains a tree-level superpotential
\be \label{Wmagadsas}
   \Wmag =\Tr \tilde X^{k+1}+\Tr\tilde X\tilde Y\tilde{\bar Y}
          +\sum_{j=0}^{k-1} \Big[M_{k-1-j} q\tilde{\bar Y}\tilde Y\bar q_{(j)}
          +N_{k-1-j} q\bar q_{(j)}+P_{k-1-j} q\tilde{\bar Y} q_{(j)}
                    +\bar P_{k-1-j}\bar q\tilde Y\bar q_{(j)}\Big]\ ,
\ee
where $q_{(j)}=\tilde X^j q$.

The electric baryons of (\ref{mesbaradsas}) can be consistently mapped to 
similar baryons of the magnetic theory:
\bea \label{barmapadsas}
   \bar\cB_p^{(n_0,\ldots,n_{k-1},\bar n_0,\ldots,\bar n_{k-1},
         \bar{\bar n}_0,\ldots,\bar{\bar n}_{k-1})} 
   &\lra 
   &\widetilde{\bar\cB}_q{}^{(m_0,\ldots,m_{k-1},\bar m_0,\ldots,\bar m_{k-1},
       \bar{\bar m}_0,\ldots,\bar{\bar m}_{k-1})}, \quad\w\nonumber\\
   q=k-p,\ m_j=N_f-\bar{\bar n}_{k-1-j}, &&\bar m_j=N_f+8-\bar n_{k-1-j},\ 
     \bar{\bar m}_j=N_f-n_{k-1-j},\nonumber\\ 
   B_n^{(n_0,\ldots,n_{k-1})} &\lra 
   &\widetilde B_m^{(m_0,\ldots,m_{k-1})},\quad\w \\
   m=k(N_f+2)-n, &&m_j=N_f+8-n_{k-1-j},\nonumber\\
   \bar B_n^{(n_0,\ldots,n_{k-1})} &\lra 
   &\widetilde{\bar B}_m^{(m_0,\ldots,m_{k-1})},\quad\w \nonumber\\
   m=2k(N_f+6)-n, &&m_j=N_f-n_{k-1-j},\nonumber
\eea
where the magnetic baryons $\widetilde{\bar\cB}_q{}^{(\ldots)}$,
$\widetilde B_m^{(\ldots)}$, $\widetilde{\bar B}_m{}^{(\ldots)}$
are defined in the same way as the electric baryons of (\ref{mesbaradsas}) 
replacing all fields by their dual partners and $N_c$ by $\tilde N_c$. 
The last two of these mappings have been found in \cite{patterns}.

For $N_c=3k(N_f+4)-1$ the magnetic theory is completely higgsed and the 
electric theory confines with low-energy spectrum given by 
the composite fields 
\bea \label{confspecadsas}
&&M_j,\ N_j,\ P_j,\  \bar P_j,\quad j=0,\ldots,k-1,\nonumber\\
&&B\equiv B_{k(N_f+2)}^{(N_f+8,\ldots,N_f+8,N_f+7)},\ 
  \bar B\equiv \cB_k^{(N_f,\ldots,N_f,N_f+8,\ldots,N_f+8,
                                      N_f,\ldots,N_f,N_f-1)},\\ 
&&\bar b\equiv \bar B_{2k(N_f+6)-1}^{(N_f,\ldots,N_f)}\nonumber
\eea 
of eqs.\ (\ref{mesbaradsas}).

The effective low-energy superpotential of the magnetic theory, deduced
from (\ref{Wmagadsas}) for the theory with $\tilde N_c=3k+1$, $(N_f+9)$ quarks
and $(N_f+1)$ antiquarks by adding a tree-level term $mM_{0}$ and integrating
out the massive modes, contains the terms 
$N_{k-1}q\bar q\ +\ P_{k-1} q\tilde{\bar Y} q$. We thus expect that 
the confining superpotential of the electric theory has terms proportional to
\bea  \label{conftermsadsas}
     \bar B N_{k-1} B &\sim 
     &(W_\alpha)^{2k}Q^{2k(N_f+8)}\bar Q^{2kN_f}X^{(k-1)(2k(N_f+4)+k-1)}
       Y^{2k(N_f+1)}\bar Y^{2k(N_f+5)},\nonumber\\
     BBP_{k-1}\bar b &\sim 
     &Q^{2k(N_f+8)}\bar Q^{2kN_f}X^{(k-1)(2k(N_f+4)-1)}
       Y^{2k(N_f+2)}\bar Y^{2k(N_f+6)}.
\eea
Comparing this to (\ref{Weff2}) we find $\alpha^{(1)}=2k$ for the term in the
first line of (\ref{conftermsadsas}) and from (\ref{albega2}, \ref{albega3}) 
$\gamma^{(1)}_1=-4k(N_f+4)+k+1$ and $\gamma^{(1)}_2=-2k(3k-1)(N_f+4)$. 
For the term in the second line we get $\alpha^{(2)}=2k$ and 
$\gamma^{(2)}_1=-4k(N_f+4)+1$ and $\gamma^{(2)}_2=-2k(3k-1)(N_f+4)+2k$. 
The confining superpotential consequently is of the form
\be \label{Wconfadsas}
    W={\bar B N_{k-1} B\ +\ h_1^{-k}h_2^{2k}BBP_{k-1}\bar b\over 
      h_1^{4k(N_f+4)-k-1}h_2^{2k(3k-1)(N_f+4)}\,
      \Lambda^{2k((3k-1)(N_f+4)-1)}}\ +\ \ldots,
\ee
where the dots denote further terms generated by instantons in the completely
broken magnetic gauge group.

In the limit $k=1$ the coupling parameter $h_1$ represents a mass for the
tensor $X$. Integrating out $X$ results in a theory with tree-level
superpotential $\Wtree=h\,\Tr(Y\bar Y)^2$, where $h=(h_2)^2/h_1$. This is 
the model discussed in section 3.3 for the special case $k=0$.
For $N_c=3(N_f+4)-1$ it confines. The  confining superpotential
(\ref{Wconfadsas}) has the correct form 
$W=(\bar B N_0 B+hBBP_0\bar b+\ldots)/h^{2(N_f+4)}\Lambda_L^{2(5(N_f+4)-2)}$
in this limit, where we used the scale matching relation 
$h_1^{N_c}\Lambda^{N_c-N_f}=\Lambda_L^{2N_c-N_f}$.

\subsection{$Sp(2N_c)$ with two \asym tensors}
Consider an $Sp(2N_c)$ gauge theory with $2N_f$ quarks $Q$ in the fundamental
representation and two traceless \asym tensors $X$, $Y$ and tree-level 
superpotential $\Wtree=h_1\,\Tr X^{k+1}\ +\ h_2\Tr XY^2$. 
This model was first studied in \cite{patterns}. 
The transformation properties of the matter fields under the 
gauge symmetry and the \nonan global symmetries are shown in the 
following table:
$$\ba{|c|c|cccc|}
\hline    \rule[-1.3ex]{0mm}{4ex}
& Sp(2N_c) & SU(2N_f) & U(1)_R & {\bfm Z}_{2(k+1)N_f} & {\bfm Z}_{2N_f} \\ 
\hline\rule{0mm}{3ex}
Q  &\Yfun &\Yfun &1-{N_c+2k+1\over(k+1)N_f} &-(N_c-1) &-(N_c-1)\\
X  &\Yasym &\bfm 1 &{2\over k+1} &2N_f &0\\
\rule[-1.5ex]{0mm}{1.5ex}
Y  &\Yasym &\bfm 1 &{k\over k+1} &-N_f &N_f\\ \hline
\ea$$

There are no baryons in symplectic gauge theories and therefore the only
(non-redundant) gauge invariant composite operators that can be built from 
the elementary fields are:
\bea \label{messp2as}
       && M_{(j,l)}=QX^jY^lQ,\quad j=0,\ldots,k-1,\ l=0,1,2,\nonumber\\
       && T_{(j,l)}=\Tr X^jY^l,
\eea
where the gauge indices are contracted with the $Sp(2N_c)$-invariant
$J$-tensor.

For odd $k$ this theory is dual to an $Sp(2\tilde N_c)$ gauge theory, with 
$\tilde N_c=3kN_f-4k-2-N_c$ and matter content \cite{patterns}
$$\ba{|c|c|cccc|}
\hline    \rule[-1.3ex]{0mm}{4ex}
& Sp(2\tilde N_c) &SU(2N_f) &U(1)_R &{\bfm Z}_{2(k+1)N_f} &{\bfm Z}_{2N_f} \\ 
\hline\rule{0mm}{3ex}
q  &\Yfun &\Yfunb &1-{\tilde N_c+2k+1\over(k+1)N_f} &N_c-3kN_f-1 &N_c-3kN_f-1\\
\tilde X  &\Yasym &\bfm 1 &{2\over k+1} &2N_f &0\\ 
\rule[-1.5ex]{0mm}{1.5ex}
\tilde Y  &\Yasym &\bfm 1 &{k\over k+1} &-N_f &N_f\\ \hline
\ea$$
and singlets $M_{(j,l)}$ that carry the same quantum numbers as the mesons 
of the electric theory.

The following tree-level superpotential of the magnetic theory is invariant
under all the symmetries:
\be \label{Wmagsp2as}
   \Wmag =\Tr\tilde X^{k+1}+\Tr\tilde X\tilde Y^2
          +\sum_{l=0}^{2}\sum_{j=0}^{k-1} M_{(k-1-j,2-l)} 
                                          q\tilde X^j\tilde Y^lq.
\ee

For $N_c=3kN_f-4k-2$ the magnetic theory is completely higgsed and the 
electric theory confines with low-energy spectrum given by the composite 
fields 
\be \label{confspecsp2as}
    M_{(j,l)},\quad j=0,\ldots,k-1,\ l=0,1,2,\quad T_{(0,2)},\ T_{(1,1)}
\ee
of eqs.\ (\ref{messp2as}). It is easy to see that the 't Hooft anomaly matching
conditions are satisfied. The contribution of $X$ and $Y$ to the global 
anomalies is $(\tilde N_c(2\tilde N_c-1)-1)$ times their fermionic charge 
under the considered symmetry. For $\tilde N_c=0$ they act therefore like 
two fields with the negative of the charge of $X$ and $Y$ respectively. 
Thus, we have to search for two composite fields of the electric theory 
that carry fermionic charge of the same absolute value as $X$ and $Y$ but 
of the opposite sign. This condition is satisfied by $T_{(0,2)}$ and 
$T_{(1,1)}$ respectively. The only other contribution to the global anomalies 
in the magnetic theory for $\tilde N_c=0$ comes from the meson singlets which 
carry the same charges as the mesons of the electric theory.

In the limit $k=1$ the coupling parameter $h_1$ represents a mass for the
tensor $X$. Integrating out $X$ results in a theory with tree-level
superpotential $\Wtree=h\,\Tr Y^4$, where $h=(h_2)^2/h_1$. This is 
the model discussed in section 3.5 for the special case $k=3$.
For $N_c=3N_f-6$ it confines. The confined spectrum (\ref{confspecsp2as})
reduces to the low-energy spectrum (\ref{confspecspas}) in this limit:
$M_{l}=M_{(0,l)}$, $l=0,1,2$, $T_3=T_{(1,1)}$. The operator $T_{(0,2)}$
gets massive\footnote{It has R-charge 1 and is neutral under all other
symmetries. Therefore a mass term is possible.} and is removed from the
low-energy spectrum. To reproduce the correct superpotential 
(\ref{Wconfspas}) $W=\sum_{\{j_m\},p}(T_3)^p\prod_{m=1}^{N_f}M_{j_m}/
h^{N_f-3}\Lambda^{5N_f-8}$ the confining superpotential of the theory
with two \asym tensors should have the expression 
$h_1^{2N_f-4}h_2^{2N_f-6}\Lambda^{2N_f-1}$ in the denominator for
$k=1$. We deduce $\gamma_1+\gamma_2=-4N_f+10$ in this limit and
obtain $\alpha=1$ from (\ref{albega}). We suppose that in general
$\alpha=k$ although we were not able to prove this.

\subsection{$Sp(2N_c)$ with an \asym and a symmetric tensor}
Consider an $Sp(2N_c)$ gauge theory with $2N_f$ quarks $Q$ in the fundamental
representation, a traceless \asym tensor $X$ and a symmetric tensor $Y$
and tree-level superpotential $\Wtree=h_1\,\Tr X^{k+1}\ +\ h_2\Tr XY^2$. 
This model was first studied in \cite{patterns}. 
The transformation properties of the matter fields under the gauge symmetry 
and the \nonan global symmetries are shown in the following table:
$$\ba{|c|c|cccc|}
\hline    \rule[-1.3ex]{0mm}{4ex}
& Sp(2N_c) & SU(2N_f) & U(1)_R & {\bfm Z}_{2(k+1)N_f} & {\bfm Z}_{2N_f} \\ 
\hline\rule{0mm}{3ex}
Q  &\Yfun &\Yfun &1-{N_c+2k-1\over(k+1)N_f} &-(N_c-3) &-(N_c+1)\\
X  &\Yasym &\bfm 1 &{2\over k+1} &2N_f &0\\
\rule[-1.5ex]{0mm}{1.5ex}
Y  &\Ysym &\bfm 1 &{k\over k+1} &-N_f &N_f\\ \hline
\ea$$

The (non-redundant) gauge invariant composite operators that can be built from
the elementary fields are:
\bea \label{messpassy}
       && M_{(j,l)}=QX^jY^lQ,\quad j=0,\ldots,k-1,\ l=0,1,2,\nonumber\\
       && T_{(j,l)}=\Tr X^jY^l,
\eea
where the gauge indices are contracted with the $Sp(2N_c)$-invariant
$J$-tensor.

For odd $k$ this theory is dual to an $Sp(2\tilde N_c)$ gauge theory, with 
$\tilde N_c=3kN_f-4k+2-N_c$ and matter content \cite{patterns}
$$\ba{|c|c|cccc|}
\hline    \rule[-1.3ex]{0mm}{4ex}
& Sp(2\tilde N_c) &SU(2N_f) &U(1)_R &{\bfm Z}_{2(k+1)N_f} &{\bfm Z}_{2N_f} \\ 
\hline\rule{0mm}{3ex}
q  &\Yfun &\Yfunb &1-{\tilde N_c+2k-1\over(k+1)N_f} 
                  &N_c-3kN_f-3 &N_c-3kN_f+1\\
\tilde X  &\Yasym &\bfm 1 &{2\over k+1} &2N_f &0\\
\rule[-1.5ex]{0mm}{1.5ex}
\tilde Y  &\Ysym &\bfm 1 &{k\over k+1} &-N_f &N_f\\ \hline
\ea$$
and singlets $M_{(j,l)}$ that carry the same quantum numbers as the mesons 
of the electric theory.

The following tree-level superpotential of the magnetic theory is invariant
under all the symmetries:
\be \label{Wmagspassy}
   \Wmag =\Tr\tilde X^{k+1}+\Tr\tilde X\tilde Y^2
          +\sum_{l=0}^{2}\sum_{j=0}^{k-1} M_{(k-1-j,2-l)} 
                                          q\tilde X^j\tilde Y^lq.
\ee

For $N_c=3kN_f-4k+2$ the magnetic theory is completely higgsed and the 
electric theory confines with low-energy spectrum given by the composite 
fields 
\be \label{confspecspassy}
    M_{(j,l)},\quad j=0,\ldots,k-1,\ l=0,1,2,\quad T_{(0,2)}
\ee
of eqs.\ (\ref{messpassy}). The argument of the previous section can be
repeated to see that the 't Hooft anomaly matching conditions are satisfied.

In the limit $k=1$ the coupling parameter $h_1$ represents a mass for the
tensor $X$. Integrating out $X$ results in a theory with tree-level
superpotential $\Wtree=h\,\Tr Y^4$, where $h=(h_2)^2/h_1$. This is 
the model discussed in section 3.4 for the special case $k=1$.
For $N_c=3N_f-2$ it confines. The confined spectrum (\ref{confspecspassy})
reduces to the low-energy spectrum (\ref{confspecspsy}) in this limit:
$M_{l}=M_{(0,l)}$, $l=0,1,2$. The operator $T_{(0,2)}$
gets massive and is removed from the low-energy spectrum. To reproduce the 
correct superpotential (\ref{Wconfspsy}) 
$W=\sum_{\{j_m\}}\prod_{m=1}^{3N_f}M_{j_m}/h^{3N_f-1}\Lambda^{3(5N_f-2)}$ 
the confining superpotential of the theory with an \asym and a symmetric 
tensor should have the expression 
$h_1^{6N_f-8}h_2^{6N_f-2}\Lambda^{3(2N_f+1)}$ in the denominator for
$k=1$. We deduce $\gamma_1+\gamma_2=-12N_f+10$ in this limit and
obtain $\alpha=3$ from (\ref{albega}). We suppose that in general
$\alpha=3k$ although we were not able to prove this.

\subsection{$SO(N_c)$ with two symmetric tensors}
Consider an $SO(N_c)$ gauge theory with $N_f$ quarks $Q$ in the fundamental
representation and two traceless symmetric tensors $X$, $Y$ and tree-level 
superpotential $\Wtree=h_1\,\Tr X^{k+1}\ +\ h_2\Tr XY^2$. 
This model was first studied in \cite{patterns}. 
The transformation properties of the matter fields under the gauge symmetry 
and the \nonan global symmetries are shown in the following table:
$$\ba{|c|c|ccccc|}
\hline    \rule[-1.3ex]{0mm}{4ex}
& SO(N_c) & SU(N_f) & U(1)_R & {\bfm Z}_{2(k+1)N_f} & {\bfm Z}_{2N_f} 
          & {\bfm Z}^\prime_{2N_f} \\ \hline
\rule{0mm}{3ex}
Q  &\Yfun &\Yfun &1-{N_c-4k-2\over(k+1)N_f} &-(N_c+2) &-(N_c+2) &1\\
X  &\Ysym &\bfm 1 &{2\over k+1} &2N_f &0 &0\\
\rule[-1.5ex]{0mm}{1.5ex}
Y  &\Ysym &\bfm 1 &{k\over k+1} &-N_f &N_f &0\\ \hline
\ea$$

We will be interested in the following gauge invariant composite operators 
that can be built from the
elementary fields:
\bea \label{mesbarso2sy}
mesons   && M_{(j,l)}=QQ_{(j,l)},\quad\w\ Q_{(j,l)}=X^jY^lQ,
            \quad j=0,\ldots,k-1,\ l=0,1,2,\nonumber\\
baryons  && \cB_p^{(n_{(0,0)},\ldots,n_{(k-1,2)})} 
           =(X^{k-1\over2}YW_\alpha)^2(YW_\alpha)^4(YXW_\alpha)^4\cdots
            (YX^{p-1}W_\alpha)^4\nonumber\\ 
         && \hskip4cm \cdot Q^{n_{(0,0)}} Q_{(1,0)}^{n_{(1,0)}}\cdots 
            Q_{(k-1,2)}^{n_{(k-1,2)}},\\
        && \qquad\w \sum_{l=0}^2\sum_{j=0}^{k-1}n_{(j,l)}=N_c-8p-4,\quad
            p=0,\ldots,\min(k,\textstyle\left\lceil{N_c-4\over8}\right\rceil),
        \nonumber
\eea
where the gauge indices are contracted with a Kronecker delta for the 
mesons and with an epsilon tensor for the baryons and we assumed $k$ odd.

For odd $k$ this theory is dual to an $SO(\tilde N_c)$ gauge theory, with 
$\tilde N_c=3kN_f+8k+4-N_c$ and matter content \cite{patterns}
$$\ba{|c|c|ccccc|}
\hline    \rule[-1.3ex]{0mm}{4ex}
& SO(\tilde N_c) &SU(N_f) & U(1)_R & {\bfm Z}_{2(k+1)N_f} & {\bfm Z}_{2N_f}
                 & {\bfm Z}^\prime_{2N_f} \\
\hline\rule{0mm}{3ex}
q  &\Yfun &\Yfunb &1-{\tilde N_c-4k-2\over(k+1)N_f} &N_c-3kN_f+2 &N_c-3kN_f+2
                                                                 &-1\\
\tilde X  &\Ysym &\bfm 1 &{2\over k+1} &2N_f &0 &0\\ 
\rule[-1.5ex]{0mm}{1.5ex}
\tilde Y  &\Ysym &\bfm 1 &{k\over k+1} &-N_f &N_f &0\\ \hline
\ea$$
and singlets $M_{(j,l)}$ that carry the same quantum numbers as the mesons 
of the electric theory.

The following tree-level superpotential of the magnetic theory is invariant
under all the symmetries:
\be \label{Wmagso2sy}
   \Wmag =\Tr\tilde X^{k+1}+\Tr\tilde X\tilde Y^2
          +\sum_{l=0}^{2}\sum_{j=0}^{k-1} M_{(k-1-j,2-l)} 
                                          q\tilde X^j\tilde Y^lq.
\ee

Under duality the electric baryons of (\ref{mesbarso2sy}) are mapped to 
the magnetic baryons $\widetilde\cB_q^{(m_{(j,l)})}=(\tilde YW_\alpha)^4\cdots
(\tilde YX^{q-1}W_\alpha)^4 q^{m_{(0,0)}}\cdots q_{(k-1,2)}^{m_{(k-1,2)}}$
according to
\be \label{barmapso2sy}
    \cB_p^{(n_{(0,0)},\ldots,n_{(k-1,2)})}\ \lra\  
    \widetilde  \cB_q^{(m_{(0,0)},\ldots,m_{(k-1,2)})},\quad \w\ 
    q=k-p,\quad m_{(j,l)}=N_f-n_{(k-1-j,2-l)}. 
\ee

For $N_c=3kN_f+8k+3$ the magnetic theory is completely higgsed and the 
electric theory confines with low-energy spectrum given by the composite 
fields 
\bea \label{confspecso2as}
     &&M_{(j,l)},\quad j=0,\ldots,k-1,\ l=0,1,2,\nonumber\\
     &&B\equiv \cB_k^{(N_f,\ldots,N_f,N_f-1)}
\eea
of eqs.\ (\ref{mesbarso2sy}).  From (\ref{barmapso2sy}) we find the mapping 
$B\lra q$. The 't Hooft anomaly matching conditions are satisfied because 
the only fields that contribute to the global anomalies in the magnetic 
theory for $\tilde N_c=1$ are the $N_f$ quark singlets staying massless
after the symmetry breaking and the meson singlets.

The effective low-energy superpotential of the magnetic theory, deduced
from (\ref{Wmagso2sy}) for the theory with $\tilde N_c=3k+1$ and $(N_f+1)$
quarks by adding a tree-level term $mM_{(0,0)}$ and integrating out 
the massive modes, contains the term $M_{(k-1,2)}qq$. We thus expect that 
the confining superpotential of the electric theory has a term proportional to
\be  \label{conftermsso2sy}
    B M_{(k-1,2)} B\ \sim\ 
    (W_\alpha)^{8k+4}Q^{6kN_f}X^{(k-1)(3kN_f+4k+1)}Y^{2(3kN_f+4k+1)}.
\ee
Comparing this to (\ref{Weff2}) we find $\alpha=3k$ and from (\ref{albega2}, 
\ref{albega3}) $\gamma_1=-6kN_f-20k-2$ and 
$\gamma_2=-3k(3k-1)N_f-24k^2-11k+1$. 
The confining superpotential consequently is of the form
\be \label{Wconfso2sy}
    W={B M_{(k-1,2)} B\over h_1^{6kN_f+20k+2}h_2^{3k(3k-1)N_f+24k^2+11k-1}\,
      \Lambda^{3k((3k-1)N_f+8k-7)}}\ +\ \ldots.
\ee

In the limit $k=1$ the coupling parameter $h_1$ represents a mass for the
tensor $X$. Integrating out $X$ results in a theory with tree-level
superpotential $\Wtree=h\,\Tr Y^4$, where $h=(h_2)^2/h_1$. This is 
the model discussed in section 3.7 for the special case $k=3$.
For $N_c=3N_f+11$ it confines \cite{newconf}. The  confining superpotential
(\ref{Wconfso2sy}) has the correct form 
$W=B M_{(0,2)} B/ h^{3N_f+17}\Lambda_L^{3(5N_f+14)}$
in this limit, where we used the scale matching relation 
$h_1^{N_c+2}\Lambda^{3(N_c-2)-2(N_c+2)-N_f}=\Lambda_L^{3(N_c-2)-(N_c+2)-N_f}$.

\subsection{$SO(N_c)$ with a symmetric and an \asym tensor}
Consider an $SO(N_c)$ gauge theory with $N_f$ quarks $Q$ in the fundamental
representation, a traceless symmetric tensor $X$ and an \asym tensor $Y$ 
and tree-level superpotential $\Wtree=h_1\,\Tr X^{k+1}\ +\ h_2\Tr XY^2$. 
This model was first studied in \cite{patterns}. 
The transformation properties of the matter fields under the gauge symmetry 
and the \nonan global symmetries are shown in the following table:
$$\ba{|c|c|ccccc|}
\hline    \rule[-1.3ex]{0mm}{4ex}
& SO(N_c) & SU(N_f) & U(1)_R & {\bfm Z}_{(k+1)N_f} & {\bfm Z}_{2N_f} 
          & {\bfm Z}^\prime_{2N_f} \\ \hline
\rule{0mm}{3ex}
Q  &\Yfun &\Yfun &1-{N_c-4k+2\over(k+1)N_f} &-(N_c+6) &-(N_c-2) &1\\
X  &\Ysym &\bfm 1 &{2\over k+1} &2N_f &0 &0\\
\rule[-1.5ex]{0mm}{1.5ex}
Y  &\Yasym &\bfm 1 &{k\over k+1} &-N_f &N_f &0\\ \hline
\ea$$

We will be interested in the following gauge invariant composite operators 
that can be built from the
elementary fields:
\bea \label{mesbarsosyas}
mesons   && M_{(j,l)}=QY^lQ_{(j)},\quad\w\ Q_{(j)}=X^jQ,
            \quad j=0,\ldots,k-1,\ l=0,1,2,\nonumber\\
baryons  && B_n^{(n_0,\ldots,n_{k-1})}
          =Y^nQ^{n_0}Q_{(1)}^{n_1}\cdots Q_{(k-1)}^{n_{k-1}},\\
        && \qquad\w \sum_{j=0}^{k-1}n_j=N_c-2n,\quad
           n=0,\ldots,\left\lceil{N_c\over 2}\right\rceil,\nonumber
\eea
where the gauge indices are contracted with a Kronecker delta for the mesons
and with an apsilon tensor for the baryons.

For odd $k$ this theory is dual to an $SO(\tilde N_c)$ gauge theory, with 
$\tilde N_c=3kN_f+8k-4-N_c$ and matter content \cite{patterns}
$$\ba{|c|c|ccccc|}
\hline    \rule[-1.3ex]{0mm}{4ex}
& SO(\tilde N_c) &SU(N_f) & U(1)_R & {\bfm Z}_{(k+1)N_f} & {\bfm Z}_{2N_f} 
                 & {\bfm Z}^\prime_{2N_f} \\ 
\hline\rule{0mm}{3ex}
q  &\Yfun &\Yfunb &1-{\tilde N_c-4k+2\over(k+1)N_f} &N_c-3kN_f+6 &N_c-3kN_f-2 
                                                    &-1\\
\tilde X  &\Ysym &\bfm 1 &{2\over k+1} &2N_f &0 &0\\
\rule[-1.5ex]{0mm}{1.5ex}
\tilde Y  &\Yasym &\bfm 1 &{k\over k+1} &-N_f &N_f &0\\ \hline
\ea$$
and singlets $M_{(j,l)}$ that carry the same quantum numbers as the mesons 
of the electric theory.

The following tree-level superpotential of the magnetic theory is invariant
under all the symmetries:
\be \label{Wmagsosyas}
   \Wmag =\Tr\tilde X^{k+1}+\Tr\tilde X\tilde Y^2
          +\sum_{l=0}^{2}\sum_{j=0}^{k-1} M_{(k-1-j,2-l)} 
                                          q\tilde X^j\tilde Y^lq.
\ee

The electric baryons of (\ref{mesbarsosyas}) can be consistently mapped to 
similar baryons of the magnetic theory:
\be \label{barmapsosyas}
   B_n^{(n_0,\ldots,n_{k-1})}\ \lra\ \widetilde B_m^{(m_0,\ldots,m_{k-1})},
   \quad\w\ m=kN_f+4k-2-n,\ m_j=N_f-n_{k-1-j}, 
\ee
where the magnetic baryons $\widetilde B_m^{(\ldots)}$,
are defined in the same way as the electric baryons of (\ref{mesbarsosyas}) 
replacing all fields by their dual partners and $N_c$ by $\tilde N_c$. 

For $N_c=3kN_f+8k-5$ the magnetic theory is completely higgsed and the 
electric theory confines with low-energy spectrum given by the composite 
fields 
\bea \label{confspecsosyas}
     &&M_{(j,l)},\quad j=0,\ldots,k-1,\ l=0,1,2,\nonumber\\
     &&B\equiv B_{kN_f+4k-2}^{(N_f,\ldots,N_f,N_f-1)}
\eea
of eqs.\ (\ref{mesbarsosyas}). From (\ref{barmapsosyas}) we find the mapping 
$B\lra q$. The 't Hooft anomaly matching conditions are satisfied because 
the only fields that contribute to the global anomalies in the magnetic 
theory for $\tilde N_c=1$ are the $N_f$ quark singlets staying massless
after the symmetry breaking and the meson singlets.

The effective low-energy superpotential of the magnetic theory, deduced
from (\ref{Wmagsosyas}) for the theory with $\tilde N_c=3k+1$ and $(N_f+1)$
quarks by adding a tree-level term $mM_{(0,0)}$ and integrating out 
the massive modes, contains the term $M_{(k-1,2)}qq$. We thus expect that 
the confining superpotential of the electric theory has a term proportional to
\be  \label{conftermssosyas}
    B M_{(k-1,2)} B\ \sim\ Q^{2kN_f}X^{(k-1)(kN_f-1)}Y^{2(kN_f+4k-1)}.
\ee
Comparing this to (\ref{Weff2}) we find $\alpha=k$ and from (\ref{albega2}) 
$\gamma_1=-2kN_f-8k+2$ and $\gamma_2=-(3k-1)kN_f-8k^2+11k-1$. 
The confining superpotential consequently is of the form
\be \label{Wconfsosyas}
    W={B M_{(k-1,2)} B\over h_1^{2kN_f+8k-2}h_2^{(3k-1)kN_f+8k^2-11k+1}\,
      \Lambda^{k((3k-1)N_f+8k-11)}}\ +\ \ldots.
\ee

In the limit $k=1$ the coupling parameter $h_1$ represents a mass for the
tensor $X$. Integrating out $X$ results in a theory with tree-level
superpotential $\Wtree=h\,\Tr Y^4$, where $h=(h_2)^2/h_1$. This is 
the model discussed in section 3.6 for the special case $k=1$.
For $N_c=3N_f+3$ it confines \cite{newconf}. The  confining superpotential
(\ref{Wconfsosyas}) has the correct form 
$W=B M_{(0,2)} B/ h^{N_f-1}\Lambda_L^{5N_f+2}$
in this limit, where we used the scale matching relation 
$h_1^{N_c+2}\Lambda^{2(N_c-2)-(N_c+2)-N_f}=\Lambda_L^{2(N_c-2)-N_f}$.

\section{Conclusions}

We have used the \nonab duality of asymptotically free $N=1$ 
supersymmetric gauge theories discovered by Seiberg to find new
models that confine in the presence of an appropriate superpotential.
This is a very interesting application of the proposed duality because
it enables us to obtain \nonpert results for the electric theory by
a perturbative calculation in its magnetic dual. Confinement in the
electric theory can be understood from the Higgs phase of the magnetic
theory. The confining spectrum can easily be derived from the
duality mappings of gauge invariant operators. For $SU$ and $SO$ 
gauge groups one also obtains the form of the confining superpotential
by applying these mappings to the magnetic tree-level superpotential.
(To determine the full confining superpotential one has to include
instanton corrections in the completely broken gauge group.)

In this paper we have discussed fifteen gauge theory models
containing fields in the fundamental representation and in second 
rank tensor representations which possess a dual description
in the infrared when an appropriate tree-level superpotential for
the tensor fields is added. All of these models confine without
breaking of chiral symmetry when the parameters (number of 
fundamental flavors $N_f$, number of colors $N_c$ and power of
the tensors in the superpotential $k$) are tuned such that the
magnetic gauge symmetry is just completely broken, i.e.\ 
one formally has $\tilde N_c=1$ for $SU$ and $SO$ gauge groups 
and $\tilde N_c=0$ for $Sp$ gauge groups, where $\tilde N_c$ is
the number of colors of the magnetic gauge theory. Reducing $N_f$
further leads to even stronger coupling in the electric theory.
In most cases the low-energy theory with one less flavor develops 
an Affleck-Dine-Seiberg superpotential and has no stable ground state. 
This is in contrast to the s-confining models with vanishing tree-level
superpotential which generically lead to confinement with quantum modified
moduli when $N_f$ is reduced by one. However, we find two examples
for gauge theories that confine with quantum modified moduli in the
presence of a tree-level superpotential: $SU(N_c)$ with an \asym
flavor and $Sp(2N_c)$ with an \asym tensor. It is an intriguing coincidence
that for vanishing tree-level superpotential these two models are the only
known examples of gauge theories that contain tensor fields and possess
a dual description for each value of $N_c$.

The phase structure of the $A_k$ models is now almost understood. However,
a deeper understanding of why just these special models can be analyzed
using duality is still missing. As for the $D_{k+2}$ models, more open
questions remain to be answered. It is not clear why a dual description
could only be found for odd $k$. Furthermore, the stability analysis
performed for the $A_k$ models has not been repeated for the $D_{k+2}$
models.

\vspace{1cm}

\begin{center} {\bf Acknowledgements} \end{center}
I would like to thank  L.~Ib\'a\~nez for helpful discussions
and  C.~Cs\'aki and H.~Murayama for useful e-mail correspondence.
This work is supported by the TMR network of the European Union, 
ref. FMRX-CT96-0090.

\end{document}